\documentclass[twocolumn,prd,aps,superscriptaddress,preprintnumbers,tightenlines,showpacs,nofootinbib,eqsecnum,amsfonts,amsmath]{revtex4}
\pdfoutput=1

\usepackage[usenames,dvipsnames]{xcolor}
\usepackage{calc}
\usepackage{amsmath,amssymb,graphicx}
\usepackage{tensor}
\usepackage{bm}
\usepackage{times}
\usepackage[varg]{txfonts}
\usepackage{float}
\usepackage{dcolumn}

\usepackage[nolist,nohyperlinks]{acronym}
\usepackage{xspace}

\usepackage[abs]{overpic}
\usepackage{pict2e}

\usepackage[caption=false]{subfig} % prevent loading of caption package: http://tex.stackexchange.com/questions/22388/subfigures-with-revtex

\usepackage[utf8]{inputenc}

\newcommand{\phD}{\textsc{IMRPhenomD}\xspace}
\newcommand{\phX}{\textsc{IMRPhenomXAS}\xspace}

\newcommand{\phXHM}{\textsc{IMRPhenomXHM}\xspace}

\DeclareMathAlphabet{\mathcalstd}{OMS}{cmsy}{m}{n}
\DeclareMathAlphabet{\mathpzc}{OT1}{pzc}{m}{it}

\definecolor{dodgerblue}{HTML}{1E90FF}
\definecolor{viennared}{HTML}{DA0A14}

\newcommand{\UIB}{Departament de F\'isica, Universitat de les Illes Balears and Institut d'Estudis Espacials de Catalunya, 
Crta. Valldemossa km 7.5, E-07122 Palma, Spain}

\begin{document}

\preprint{LIGO-PXXX}

%%%%%%%%%%%%%%%%%%%%%%%%%%%%%%%%%%%%%%%%%%%%%%%%%%%% Title page %%%%%%%%%%%%%%%%%%%%%%%%%%%%%%%%%%%%%%%%%%%%%%%%%%%%

\title{Accelerating the evaluation of inspiral-merger-ringdown waveforms with adapted grids}

\author{Cecilio Garc{\'i}a-Quir{\'o}s}
\affiliation{\UIB}

% \author{Marta Colleoni}
% \affiliation{\UIB}

\author{Sascha Husa}
\affiliation{\UIB}
%\affiliation{\ICTS}

% \author{H{\'e}ctor Estell{\'e}s}
% \affiliation{\UIB}

% \author{Geraint Pratten}
% \affiliation{\UoB}
% \affiliation{\UIB}

%\author{Antoni Ramos-Buades}
%\affiliation{\UIB}

\author{Maite Mateu-Lucena}
\affiliation{\UIB}

\author{Angela Borchers}
\affiliation{\UIB}

% \author{Rafel Jaume}
% \affiliation{\UIB}

%\affiliation{\UIB}

%\affiliation{\UIB}

%\affiliation{\ICTS}

%%%%%%%%%%%%%%%%%%%%%%%%%%%%%%%%%%%%%%%%%%%%%%%%%%% Abstract %%%%%%%%%%%%%%%%%%%%%%%%%%%%%%%%%%%%%%%%%%%%%%%%%%%%%%
\begin{abstract}

This paper presents an algorithm to accelerate the evaluation of inspiral-merger-ringdown waveform models for gravitational wave data analysis. While the idea can also be applied in the time domain, here we focus on the frequency domain, which is most typically used to reduced computational cost in gravitational wave data analysis. 
Our work extends the idea of multibanding \cite{Vinciguerra:2017ngf}, which has been developed to accelerate frequency domain waveforms,
to include the merger and ringdown and spherical harmonics beyond the dominant quadrupole spherical harmonic.
The original method of \cite{Vinciguerra:2017ngf} is based on a heuristic algorithm based on the inspiral to de-refine the equi-spaced frequency grid used for data analysis where a coarser grid is sufficient for accurate evaluation of a waveform model.
Here we use a different criterion, based on the local interpolation error, which is more flexible and can easily be adapted to general waveforms, if their phenomenology is understood.
We discuss our implementation in the LIGO Algorithms library \cite{lalsuite} for the PhenomXHM \cite{phenXHM} frequency domain model, and report the acceleration in different parts of the parameter space of compact binary systems.
\end{abstract}


\today

\maketitle

% ======================
%  ACRONYMS
% ======================
\acrodef{PN}{post-Newtonian}
\acrodef{EOB}{effective-one-body}
\acrodef{NR}{numerical relativity}
\acrodef{GW}{gravitational-wave}
\acrodef{BBH}{binary black hole}
\acrodef{BH}{black hole}
\acrodef{BNS}{binary neutron star}
\acrodef{NSBH}{neutron star-black hole}
\acrodef{SNR}{signal-to-noise ratio}
\acrodef{aLIGO}{Advanced LIGO}
\acrodef{AdV}{Advanced Virgo}

\newcommand{\PN}[0]{\ac{PN}\xspace}
\newcommand{\EOB}[0]{\ac{EOB}\xspace}
\newcommand{\NR}[0]{\ac{NR}\xspace}
\newcommand{\BBH}[0]{\ac{BBH}\xspace}
\newcommand{\BH}[0]{\ac{BH}\xspace}
\newcommand{\BNS}[0]{\ac{BNS}\xspace}
\newcommand{\NSBH}[0]{\ac{NSBH}\xspace}
\newcommand{\GW}[0]{\ac{GW}\xspace}
\newcommand{\SNR}[0]{\ac{SNR}\xspace}
\newcommand{\aLIGO}[0]{\ac{aLIGO}\xspace}
\newcommand{\AdV}[0]{\ac{AdV}\xspace}

%%%%%%%%%%%%%%%%%%%%%%%%%%%%%%%%%%%%%%%%%%%%%%%%%%%%%%%
\section{Introduction}\label{sec:introduction}

The field of gravitational wave astronomy has been born through discoveries of coalescences of compact binary systems consisting of black holes and neutron stars \cite{PhysRevLett.116.061102,PhysRevLett.119.161101,LIGOScientific:2018mvr}.
For such systems, very successful programs are being carried out to model the gravitational waveforms expected according to general relativity (and possibly alternative theories) across the astrophysically plausible parameter space of observable binary systems (see e.g.~ \cite{Taracchini:2014zpa,Bohe:2016gbl,Ajith2007,Ajith2011,Hannam:2013oca,Husa:2015iqa,Khan:2015jqa,Blackman:2015pia}).
These models are based on synthesising perturbative results, e.g. from post-Newtonian theory \cite{Blanchet2014}, black hole perturbation theory \cite{Kokkotas1999} and more recently the self-force approach \cite{Barack:2018yvs}, 
with numerical solutions of the Einstein equations, with an important role played by the effective-one-body approach \cite{Buonanno:1998gg,Buonanno1999} to extend perturbative to non-perturbative descriptions.

Gravitational wave data analysis as applied to compact binary coalescence is typically split into two steps: searches and Bayesian parameter estimation.
Searches can be performed independently from a waveform model \cite{Klimenko:2015ypf}, or use 
 a fixed set of template waveforms and matched filter techniques \cite{Usman:2015kfa,Sachdev:2019vvd}.
Bayesian parameter estimation \cite{Veitch:2014wba,Ashton:2018jfp} is based on a likelihood function that compares the detector data with template waveforms.  
Several million template waveform evaluations may be required, and the computational cost of waveform evaluation makes Bayesian inference computationally expensive.
In this paper we discuss the problem of accelerating the evaluation of the waveforms, intended in particular to reduce the computational cost of Bayesian parameter estimation.

A particularly computationally efficient approach to the construction of waveform models has been the phenomenological waveform approach (see e.g.~\cite{Ajith2007,Ajith2011,Hannam:2013oca,Husa:2015iqa,Khan:2015jqa}), where the waveform for each spherical harmonic is split into a small number (typically 2-4) regions based on physical intuition, and are written as closed form expressions. In order to model simple non-oscillatory functions, it is further customary to split the waveform $h_{\ell m}(x, \Xi)$ for spherical harmonic $(\ell,m)$ into a real amplitude $A_{\ell,m}(x, \Xi)$ and a phase $\phi_{\ell,m}(x, \Xi)$. Here $h$ would typically be the gravitational wave strain or its Fourier transform, and $x$ the time or frequency, respectively. The quantity  $\Xi$ is a shorthand for all the intrinsic parameters of the waveform, such as masses and spins.  
We then compute the waveform of each spherical harmonic as
\begin{equation}\label{eq:amp_phase_split}
h_{\ell m} = A_{\ell m} e^{i \, \phi_{\ell m}}. 
\end{equation}

The evaluation of matched filters (e.g. due to optimization over time of arrival) typically requires the evaluation of fast Fourier transforms, which require equispaced grids.
Typically, a computationally much cheaper interpolant could be constructed by only evaluating the model amplitude and phase on a much coarser grid without
significant loss of accuracy, if the coarse grid points are chosen judiciously.
Our goal is the same as that of \cite{Vinciguerra:2017ngf}: to accelerate the evaluation of $A_{\ell m}$  and $\phi_{\ell m}$, but also the calculation of the complex exponential $e^{i \, \phi_{\ell m}}$, through an appropriate choice of coarse grid points and interpolation algorithm. 
For simplicity we will also us use the term ``multibanding'' to refer to this type of algorithm, and
we also use the same two core ideas:
\begin{itemize}
\item We split the complete frequency or time range where we want to evaluate our model waveform into $n$ sub-regions, where each region has a constant grid spacing $\Delta x_n$, chosen such that linear interpolation is sufficiently accurate for a given criterion of waveform accuracy. The final waveform can then be evaluated by simple linear  interpolation to the fine grid with constant grid-spacing $dx$, which is determined by the requirements of gravitational wave data analysis. This step accelerates the evaluation of the amplitude and phase.
\item For the phase, the computationally expensive evaluation of the complex exponential in eq.~(\ref{eq:amp_phase_split}) for each point of the fine grid is required.
For coarse grids that are sufficiently dense for linear interpolation, a standard algorithm can be used to replace evaluation of the complex exponential at each point of the fine grid by evaluation only at the coarse grid points, and implementing linear interpolation as an iterative scheme.
\end{itemize}

The key difference between our work and \cite{Vinciguerra:2017ngf}
is that we change the criterion to compute the grid spacings $\Delta x_n$ in the $n$ coarse grids to use the standard estimate of the local interpolation error derived according to Taylor's theorem of basic calculus instead of a heuristic algorithm based on the relation between the duration of a data segment and the frequency spacing in the Fourier domain. Below we will analyze the required frequency spacing for the inspiral, merger and ringdown. We will first carry out the analysis separately for the amplitude and phase of different modes, and then define coarse grids that are appropriate for both phase and amplitude for each mode. For an overview of how the interpolation is incorporated in the context of the Reduced Order Models (ROM) see \cite{P_rrer_2014}.

This paper is organized as follows: In Sec.~\ref{sec:algorithm} we discuss the details of this algorithm, and how it is applied to quasi-circular non-precessing frequency domain waveforms
for the inspiral, merger and ringdown.  In Sec.~\ref{sec:results} we present our results for computational efficiency and accuracy, and we conclude with a summary and comments on possible future work in 
Sec.~\ref{sec:conclusions}.

%%%%%%%%%%%%%%%%%%%%%%%%%%%%%%%%%%%%%%%%%%%%%%%%%%%%%%%%
\section{Algorithms} \label{sec:algorithm}

\subsection{Interpolation error} \label{sec:interp} 

A real-valued differentiable function $g(x)$ can be approximated at a point $x_0$ by a linear approximation in the following sense:
There exists a function $h(x)$ such that
\begin{equation}
g(x)=g(x_0)+g'(x_0)(x-x_0)+h(x)(x-x_0),\quad \lim _{x\to x_0}h(x)=0.
\end{equation}
The error $R(x)$ of the approximation is
\begin{equation}
R(x)=h(x)(x-x_0).
\end{equation}

According to standard refinements of Taylor's theorem of basic calculus, the error term $R(x)$ can be estimated using the second derivative $g''(x)$ of the function $g$ we want to approximate by the statement that there exists a $\xi$,  $x_0 \leq \xi \leq x$, such that
\begin{equation}
R(x) =  \frac{g''(\xi)}{2} \left( x- x_0 \right)^2.
\end{equation}
If we apply this result to our problem of interpolating to a fine grid from a coarse grid with grid spacing $\Delta x_n$,
then
\begin{equation}
R(x) \leq \max_{x_0 \leq \xi \leq x}\frac{g''(\xi)}{2} {\Delta x_n}^2.
\end{equation}
Consequently we can choose our coarse grid spacing $\Delta x_n$ to satisfy a given error threshold $R$
as 
\begin{equation}\label{eq:Deltax_of_R}
\Delta x_n = \sqrt{\frac{2 R}{ \max_{x_0 \leq \xi \leq x} g''(\xi)} }.
\end{equation}

Our application of interpolation will initially be guided by the requirements of phase accuracy, and we will then discuss
in which sense these criteria also lead to a sufficiently small amplitude error. Below we will develop the details of constructing a hierarchy of grids as appropriate for linear interpolation of both the frequency domain phase
and amplitude for different spherical or spheroidal harmonic modes, and describe how to efficiently evaluate complex exponentials of the phase on such a grid hierarchy. 
%
%For simplicity, we will here use the same grid hierarchy for the amplitude, we will however take the freedom to evaluate it with third
%order interpolation on the fine grid at little extra cost
%-- for the amplitude no complications since we do not have to evaluate a complex exponential.

\begin{figure}[ht]
    \centering
    \includegraphics[width=\columnwidth]{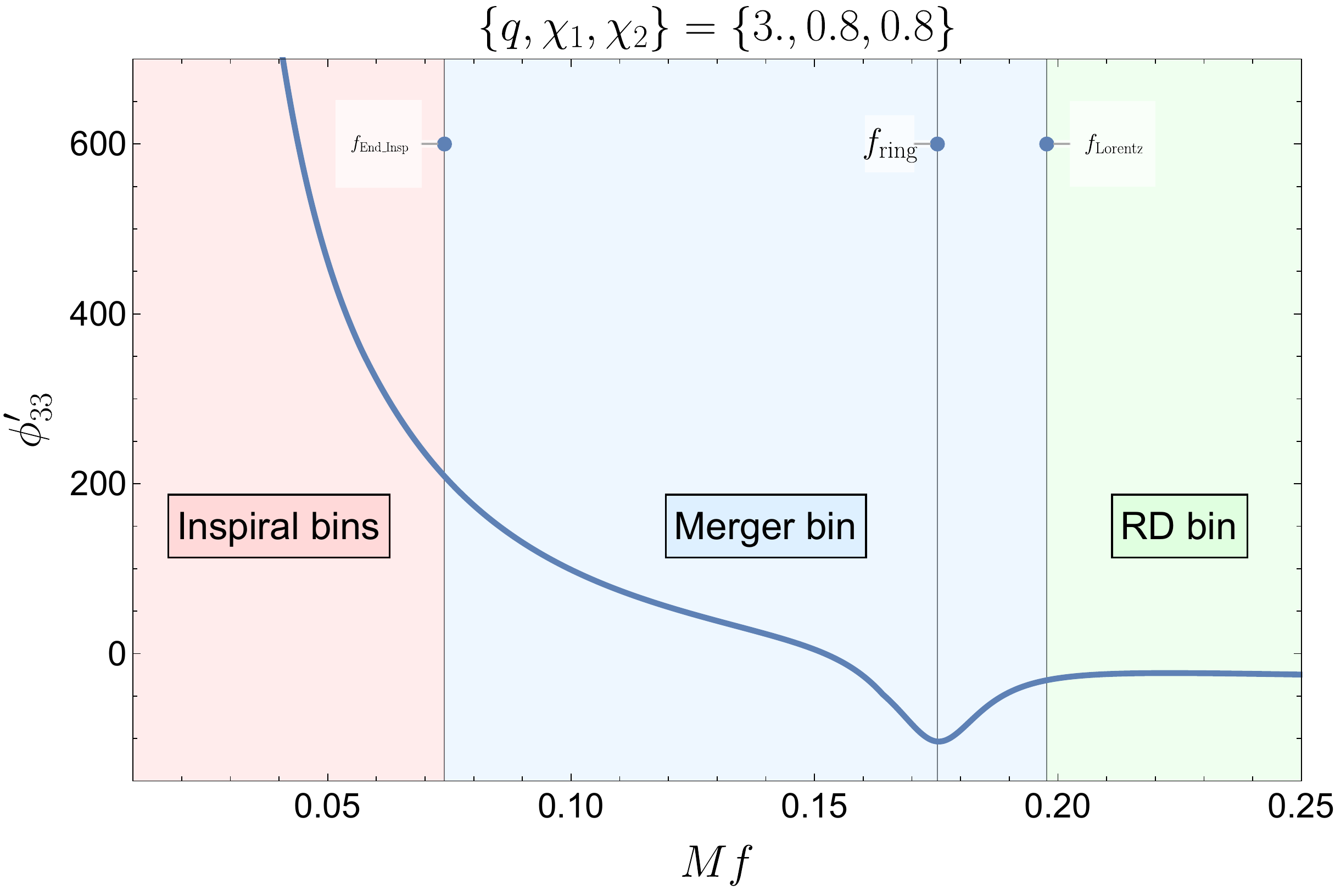}
     \includegraphics[width=\columnwidth]{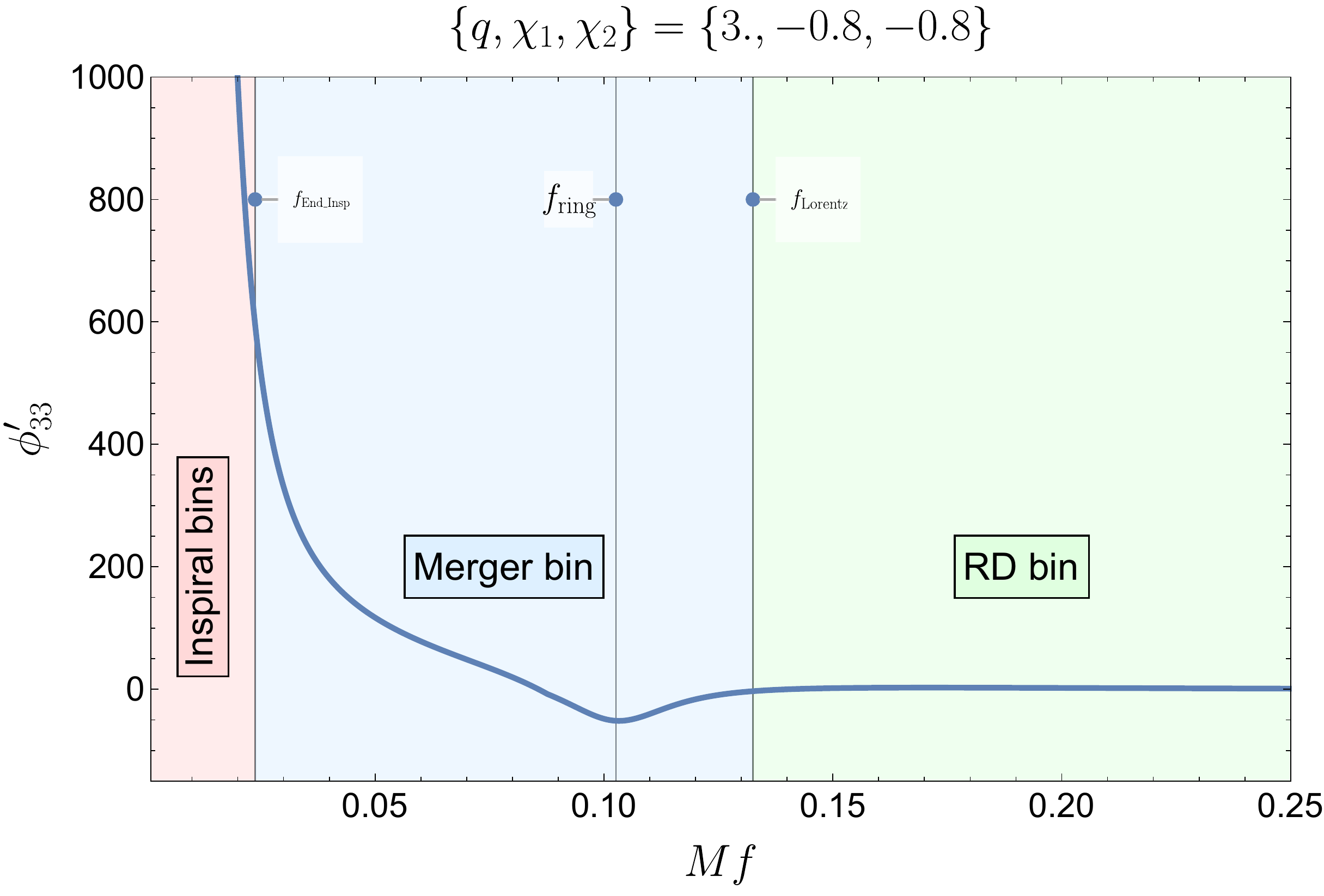}
    \caption{The phase derivative is shown for the $\ell=\vert m\vert = 3$ spherical harmonic mode for two different configurations with high spins. The phase derivative changes rapidly in the inspiral region, and thus many grid points are required for accurate interpolation, the ringdown part is however comparatively flat, thus only few points are needed to describe it. The merger bin is characterized by the shape of the Lorentzian and determines the resolution required in this region.}
    \label{fig:derivative}
\end{figure}

The hierarchy of grids is determined by the behaviour of the second derivative of the phase as a function of the frequency according to eq.~(\ref{eq:Deltax_of_R}). We distinguish between three main regions: inspiral, merger and ringdown. As shown in Fig.~\ref{fig:derivative} the behaviour of the phase derivative is sharper and changes very drastically, then more points will be needed in this region. However the merger and ringdown parts are ``flatter'' and less points will be necessary to describe these parts.

\subsubsection{Inspiral in the frequency domain} \label{sec:insp_freq} 
%%%%%%%%%%%%%%%%%%%%%%%%%%%%%%%%%%%%%%%%%%%%%%%%%%%%%%%%%%%%%%%%%%%%%%

In order to derive an appropriate frequency grid spacing $\Delta f$ ($f$ is the dimensionless frequency in geometric units $G=c=1$) for the Fourier domain phase during inspiral, we will approximate the phase by the leading TaylorF2 phase expression \cite{Buonanno:2009zt},
\begin{equation}
%\Phi_{\ell m} = c_0 + c_1 f + \frac{m}{2} \frac{3}{128 \eta} (\pi f)^{-5/3},
\Phi_{\ell m} = c_0 + c_1 f + \frac{m}{2} \frac{3}{128 \eta} \left(\frac{2\pi f}{m}\right)^{-5/3},
\end{equation}
where $\eta$  is the symmetric mass ratio, which in terms of the component masses $m_{1,2}$ of the binary reads $\eta = m_1 m_2/(m_1 + m_2)^2$,
and $c_0$, $c_1$ are constants of integration that do not affect the second derivative, which reads
\begin{equation}
%\Phi''_{\ell m} = \frac{m}{2}  \frac{5}{48 \eta} \pi^{-5/3} f^{-11/3}.
\Phi''_{\ell m} = \frac{m}{2}  \frac{5}{48 \eta} \left(\frac{2\pi}{m}\right)^{-5/3} f^{-11/3}.
\end{equation}
The phase and phase derivatives becomes singular as the frequency $f$ approaches zero, and the magnitude of the second derivative increases toward decreasing frequency. We can therefore estimate the maximal second phase derivative as the second derivative of the phase at the start of each frequency interval for which we want to use interpolation,
assumed at frequency $f$, and obtain
\begin{eqnarray}\label{eq:df_insp}
%\Delta f (f) &=& \sqrt{\frac{2 R}{\Phi''(f)} } =  \sqrt{2 R c_{f,insp}} f^{11/6}, \\
\Delta f (f) &=& \sqrt{\frac{2 R}{\Phi''(f)} } =  \sqrt{\frac{2 R}{c_{f,insp}}} f^{11/6}, \\
 c_{f,insp} &=&  \frac{m}{2}  \frac{5}{48 \eta} \left(\frac{2\pi}{m}\right)^{-5/3}.
\end{eqnarray}

In Sec.~\ref{sec:mb_frequ} we will use eq.~(\ref{eq:df_insp}) to split the calculation of the phase into frequency bins, where in each bin the grid spacing $\Delta f$ is kept constant, but it increases from bin to bin with increasing start frequency of the bin. 

We now turn to the inspiral amplitude. For the modes we consider in this work
we obtain the following leading order terms, see e.g.~\cite{phenXHM},
where we use the definitions $v=(2\pi f/m)^{1/3}$ and 
$\delta = \sqrt{1- 4\eta}$:
\begin{eqnarray}\label{eq:Alm}
A_{\ell m} &=& \pi \sqrt{2 \eta/3} \, v^{-7/2}  a_{\ell m},\\\label{eq:A22}
a_{22} &=& 1 + O(v^2),\\\label{eq:A21}
a_{21} &=& v \delta \frac{\sqrt{2}}{3} + O(v^2),\\\label{eq:A33}
a_{33} &=& v \delta \frac{3}{4} \sqrt{\frac{5}{7}} + O(v^3),\\\label{eq:A32}
a_{32} &=& v^2 \frac{1}{3} \sqrt{\frac{ 5}{7}}  \, (1 - 3 \eta) + O(v^3),\\\label{eq:A44}
a_{44} &=& v^2 \frac{4}{9} \sqrt{\frac{10}{7}} \, (1-3 \eta) + O(v^4). 
\end{eqnarray}

For the amplitude it is natural to define a threshold for the relative error of the interpolation, which we denote by $r$. The frequency dependent coarse grid resolution $\Delta f(f)$ which results from specifying a relative error threshold $r$ is then independent of $\eta$ and depends linearly on the frequency,
\begin{equation}
\Delta f_{\ell m}(f) = \sqrt{\frac{2r A_{lm}}{A_{lm}''}} = c_{\ell m} \, f \sqrt{r},    
\end{equation}
where
\begin{eqnarray}\label{eq:clm}
 c_{22} &=& 6 \sqrt{\frac{2}{91}},\\
 c_{21} &=& c_{33} = 6 \sqrt{\frac{2}{55}},\\
 c_{44} &=& c_{32} = 2 \sqrt{\frac{2}{3}}.
\end{eqnarray} 
We can then write the ratio of coarse grid spacing required for the phase to stay below a phase error of $R$ radians to the coarse grid spacing required for the amplitude to guarantee a relative amplitude error below $r$ as
% \begin{equation}
% \frac{\Delta f_{phase}}{\Delta f_{amp}} = \frac{\alpha_{\ell m}}{\sqrt{\eta}} \left(\frac{f}{\pi}\right)^{5/6} \sqrt{\frac{r}{R}},    
% \end{equation}
\begin{equation}
\label{eq:ratio_deltaf}
\frac{\Delta f_{phase}}{\Delta f_{amp}} = \alpha_{\ell m} \left(f\pi\right)^{5/6} \sqrt{\eta}\sqrt{\frac{R}{r}},    
\end{equation}
where
% \begin{eqnarray}
%   \alpha_{22} &=& \frac{\sqrt{\frac{455}{3}}}{24 }, \quad  
%   \alpha_{21} = \frac{5 \sqrt{\frac{11}{6}}}{24}, \\  
%   \alpha_{33} &=& \frac{5 \sqrt{\frac{11}{2}}}{24 }, \quad   
%   \alpha_{32} = \frac{\sqrt{5}}{8 }, \quad   
%   \alpha_{44} = \frac{\sqrt{\frac{5}{2}}}{4}.
% \end{eqnarray}
\begin{eqnarray}
   \alpha_{22} &=& 2 \sqrt{\frac{91}{55}}, \quad  
   \alpha_{21} = 2^{1/3}\sqrt{\frac{11}{3}}, \\  
   \alpha_{33} &=& \frac{2^{7/3}}{3^{11/6}} \sqrt{11}, \quad   
   \alpha_{32} = \frac{6}{\sqrt{5}}, \quad   
   \alpha_{44} = \frac{3}{2^{1/3}\sqrt{5}}.
\end{eqnarray}

Choosing e.g.~$r=R$ the expression (\ref{eq:ratio_deltaf}) is always smaller than unity up to the minimum energy circular orbit (MECO) frequency \cite{Cabero:2016ayq}, the step size restriction for the phase is thus more restrictive than the one for the amplitude. For simplicity we will use the phase criteria to build just one coarse frequency array and use this for both the phase and amplitude. We discuss the merger and ringdown in the next section.

\subsubsection{Merger and ringdown in the frequency domain} \label{sec:merg_freq} 
%%%%%%%%%%%%%%%%%%%%%%%%%%%%%%%%%%%%%%%%%%%%%%%%%%%%%%%%%%%%%%%%%%%%%%

The merger-ringdown phase exhibits a morphology that is rather different from the inspiral. A detailed phenomenological description for the 
$\ell=\vert m \vert = 2$ is provided by the \phD \cite{Husa:2015iqa} and
\phX \cite{phenX} waveform models, and for subdominant modes by \phXHM \cite{phenXHM}. This allows us to identify the crucial features of the merger-ringdown regime, and to adapt the estimate (\ref{eq:Deltax_of_R})
for the step size as we have done for the inspiral.

%changes significantly its behaviour compared to the inspiral part. For that reason we modify and adapt the algorithm to this particular region. In first place the spacing of the coarse frequency grid given a particular frequency is altered.

The first ingredient will be to identify the end of the inspiral. In \cite{hybrids,phenX,phenXHM} we confirm that the minimum energy circular orbit (MECO) \cite{Cabero:2016ayq} provides a good approximation for the transition between inspiral and merger for comparable masses.
In the merger-ringdown part the Fourier domain phase derivative is given by a superposition of a Lorentzian function and a background term \cite{phenX,phenXHM}. The Lorentzian dominates the phase derivative and is given by (see eq.~(6.3) in \cite{phenXHM}):
\begin{equation}\label{eq:Lorentzian}
\Phi^{\prime}(f) = \frac{a}{\left( f-f_0\right)^2 + b^2}
\end{equation}
and the second derivative by
\begin{equation}\label{eq:Der2Lorentzian}
   \Phi^{\prime \prime}(f) = -\frac{2 a (f - f_0)}{((f-f_0)^2 + b^2)^2},
\end{equation}
where we have introduced the shorthands $a = a_{\lambda} f_{\mathrm{damp}}^{lm} $, $b = f_{\mathrm{damp}}^{lm}$, and $f_0 =f_{\mathrm{ring}}^{lm}$. Thus $\alpha_{\lambda}$ is a term that determines the overall amplitude of the Lorentzian, $f_0$ is the frequency at which the dip of the Lorentzian happens and $b$ is a measure of the "width" of the dip.
Inserting the expression for the Lorentzian into eq.~(\ref{eq:Deltax_of_R}) we obtain
\begin{equation}
\label{eq:df_mrd}
\Delta f (f) = \sqrt{\frac{2 R}{\Phi''(f)} } = \sqrt{\frac{R }{a |f-f_0|} } \left( \left( f-f_0\right)^2 + b^2\right),
\end{equation}
which replaces eq.~(\ref{eq:df_insp}) for computing the spacing of the coarse frequency grid in the merger and ringdown. 

The spacing computed according to (\ref{eq:Deltax_of_R}) depends on the absolute value of the second derivative, and we note that the second derivative of the Lorentzian phase function, $\Phi''$, has two local maxima for $f_0 \pm b/\sqrt{3}$, with identical absolute value
\begin{equation}\label{eq:max_ddphi_lorentzian}
    \left\vert \Phi''\left(f_0 \pm \frac{b}{\sqrt{3}}\right) \right\vert = \frac{3 \sqrt{3} a}{8 b^3}.
\end{equation}

While for the inspiral the number of frequency bins depends on the start frequency, as we will discuss in more detail below in Sec.~\ref{sec:mb_frequ}, for the merger and ringdown we choose two bins which we call the \textit{merger} and \textit{ringdown} bins. The merger bin is defined by the frequency interval $\left( f_{\mathrm{Insp}}, f_{\mathrm{Lorentzian}}  \right)$, 
and captures the frequency regime where high resolution is required to capture the shape of the Lorentzian. The frequency $f_{\mathrm{Insp}}$ marks the end of the inspiral region of the \phX model for the $\ell = \vert m\vert = 2$ mode and the \phXHM model for the other harmonics, and is chosen approximately at the MECO frequency (see \cite{phenX} and \cite{phenXHM} for details). 
The frequency $f_{\mathrm{Lorentzian}}$ is defined as $f_{\mathrm{ring}}^{lm} + 2  f_{\mathrm{damp}}^{lm}$ and is chosen to approximate the lowest frequency where the second phase derivative of the Lorentzian can be neglected, and the first phase derivative is approximately constant. 
The ringdown bin starts at this frequency, and is the highest frequency bin in our procedure. It is characterized by low resolution requirements for the phase due to neglecting $\Phi''$ and ends at the end frequency of the waveform, the frequency range of this last bin is thus  $\left( f_{\mathrm{Lorentzian}}, f_{\mathrm{max}}  \right)$. 

We compute the grid spacing of both bins by evaluating the maximum value of $|\Phi''|$ in these two intervals according to eqs.~(\ref{eq:df_insp}, \ref{eq:max_ddphi_lorentzian}) and inserting it into eq.~(\ref{eq:Deltax_of_R}).
For the merger bin this is the maximum of the inspiral value and the  value for the Lorentzian, and thus
% \begin{equation}
% \label{eq:dfmerger}
% \Delta f_{\mathrm{merger}} = \min{ \left( 
% \sqrt{2 R c_{f,insp}} (f_{insp})^{11/6},
% \frac{4 f_{\mathrm{damp}}^{lm}}{3^{3/4}}\sqrt{\frac{R}{|\alpha_{\lambda}|}} \right) },
% \end{equation}
\begin{equation}
\label{eq:dfmerger}
\Delta f_{\mathrm{merger}}^{phase} = \min{ \left( 
\sqrt{\frac{2 R}{c_{f,insp}}} f_{insp}^{\frac{11}{6}},
\frac{4 f_{\mathrm{damp}}^{lm}}{3^{3/4}}\sqrt{\frac{R}{|\alpha_{\lambda}|}} \right) },
\end{equation}

For the ringdown bin the second phase derivative $|\Phi''|$ decreases monotonically to zero, we thus take the value at the start of the region $f_{\mathrm{Lorentzian}}$, which yields
% \begin{equation}
%     \Phi''(f_{\mathrm{Lorentzian}}) = \frac{4 a}{25 f_{\mathrm{damp}}^{lm}}.
% \end{equation}
%Inserting these two frequencies in (\ref{eq:Der2Lorentzia}) and performing we get
%
% \begin{equation}
% \label{eq:dfringdown}
% \Delta f_{\mathrm{RD}} = 5 \sqrt{\frac{R (f_{\mathrm{damp}}^{lm})^3}{|\alpha_{\lambda}|}}.
% \end{equation}
\begin{equation}
\label{eq:dfringdown}
\Delta f_{\mathrm{RD}}^{phase} = 5 \:f_{\mathrm{damp}}^{lm} \sqrt{\frac{R}{2|\alpha_{\lambda}|}}.
\end{equation}

Again we turn to the amplitude now. We approximate the amplitude falloff in the ringdown bin as
\begin{equation}
    h \approx e^{-\Lambda f},
\end{equation}
with $\Lambda = \lambda / (f_{damp}^{lm} \sigma)$, where these coefficients correspond to those used in the ringdown ansatz for the \phXHM model (see \cite{phenXHM} for more details):
\begin{equation} \label{eq:ringdown_ansatz}
A_{RD}^{\ell m} \propto \frac{1}{\left(f-f^{\ell m}_{ring}\right)^2+\left(f^{\ell m}_{damp}\:\sigma\right)^2}e^{-\frac{ \left(f-f^{\ell m}_{ring}\right)\:\lambda}{f^{\ell m}_{damp}\sigma}}.
\end{equation}
The grid spacing required to guarantee a relative error smaller than $r$ is then given by
\begin{equation}\label{eq:df_amp_ringdown}
   \Delta f_{RD}^{amp}(r) =\frac{\sqrt{2 \, r}}{\Lambda },
\end{equation}
which is independent of the frequency $f$.
For $r=R$ this condition is typically more restrictive than the condition (\ref{eq:dfringdown}) derived from the phase, the dependence across parameter space is however complicated. We therefore always compute the two frequency spacings, and then use the more restrictive one. We believe that this choice is quite conservative and that the choice could be relaxed in the future, since our ringdown bin only starts at frequencies where the amplitude is already quite small. Note that the start frequency of our ringdown region is either significantly higher than the ringdown frequency, or, for very high spins, the exponential falloff is significantly steeper than for moderate spins. In consequence we could use always the phase criterion  (\ref{eq:dfmerger}) to set the grid spacing in the ringdown region without worrying too much about loss of accuracy. If greater amplitude accuracy for the ringdown would be required, it would be also possible to switch from linear interpolation to the fine grid to third order spline interpolation for the amplitude.
%\sascha{For $r=R$ this condition is typically more restrictive than the condition (\ref{eq:dfmerger}) derived from the phase. Our ringdown region however only starts at frequencies where the amplitude is already quite small -- note that the start frequency of our ringdown region is either significantly higher than the ringdown frequency, or, for very high spins, the exponential falloff is significantly steeper than for moderate spins. In consequence our current implementation uses the phase criterion  (\ref{eq:dfmerger}) to set the grid spacing in the ringdown region. If greater amplitude accuracy for the ringdown is required, it is possible to switch from linear interpolation to the fine grid to third order spline interpolation for the amplitude.}

In the merger bin, the functional dependence of the mode amplitudes is more complex (see \cite{phenXHM}). In this case we compute numerically the grid spacing $\Delta f$ for the amplitude as 
$$
\Delta f_{merger}^{amp} = \sqrt{\frac{2 \:r \: |h_{lm}(f)|}{|h_{lm}(f)|''}}.
$$
We evaluate this quantity for the merger bin across our parameter space with the choice $r=R$ and compare with the grid spacing derived for the phase given by eq.~(\ref{eq:dfmerger}). We find that the ratio 
$\Delta f_{merger}^{phase}/\Delta f_{merger}^{amp}$ is typically lower than one so the criteria for the phase is more restrictive than the one for the amplitude. We find that for some cases with comparable masses and high positive spins the ratio is between a value of one and two, but for simplicity we will  always choose the criterion for the phase and interpret this choice such that the actual relative amplitude and phase errors will be bounded by the thresholds within a factor of four. We leave refinements of the simple strategy to set
$r = R$ to future work. Below we will study the mismatch between the original model and different levels of error threshold to arrive at a more practical evaluation of error than to check for local deviations between model and approximation, and in Sec.~\ref{sec:pe} we will perform a parameter estimation exercise and find that all choices of the value of $R=(0.1, 0.01, 0.001)$ lead to indistinguishable results for the case considered.

%\sascha{We evaluate this quantity for the merger bin across our parameter space and compare with the grid spacing derived for the phase given by eq.~\ref{eq:dfmerger}), and we find that the grid spacing inferred from the phase guarantees that the relative error in the amplitude $r$ agrees with the absolute error in the phase $R$ within roughly a factor of two.}

% \subsubsection{Inspiral in the time domain} \label{sec:insp_time} 

% Now we make the intervals smaller toward the merger.

% \subsubsection{Merger and ringdown in the time domain} \label{sec:merg_time} 

% Can we do something smart, or we do something very naive like currently for the frequency domain?

\subsection{Efficient evaluation of complex exponentials} \label{sec:exp_eval} 

The evaluation of the complex exponential function when constructing the strain
from amplitude and phase as in eq.~(\ref{eq:amp_phase_split})
is one of the most time consuming operations in the C-code of the \texttt{LALSuite} \cite{lalsuite} implementation of our model. The number of required evaluations
of the complex exponential (or, equivalently, of trigonometric functions), can
however be reduced drastically by implementing the method described in \cite{Vinciguerra:2017ngf} (adapted from \cite{numRecipes}). Instead of interpolating the phase on the uniform fine grid and computing the complex exponential, we compute the complex exponential in the non-uniform coarse grid and then rewrite the interpolation of this quantity in terms of an iterative algorithm.

Let $\Phi_j$ be the phase at one coarse frequency point $f_j$ and let $\hat{\Phi}_k$, $f_k$ be the estimated phase and the frequency at one point of the final uniform frequency grid, the spacing of the uniform grid is therefore $df=f_{k+1}-f_{k}$. Then we use the recursive property
\begin{equation}
    e^{i \hat{\Phi}_{k+1}} = e^{i \hat{\Phi}_k} e^{i\:df \: \frac{\Phi_{j+1} - \Phi_j}{f_{j+1}-f_j}}. 
\end{equation}
This property is used to compute the complex exponential in the fine frequency grid points that lay between two coarse frequency points $j$ and $j+1$. The first of the fine points is given by  $e^{i \hat{\Phi}_0} =  e^{i \Phi_j}$.

\subsection{Complete multibanding algorithm in the frequency domain} \label{sec:mb_frequ} 

We will now describe our final algorithm for accelerated waveform evaluation, which is based on our previous results. Our final results will be the strain,
evaluated on a uniform frequency grid, with a resolution $df$ that is adapted to the requirements of some given data analysis application. The motivation for uniform grid spacing stems for the typical context of matched filtering, where an inverse Fourier transform is used to optimize a match over the time shift between a signal and a template. We will refer to this uniform frequency grid as the fine grid. In order to accelerate the waveform evaluation we will however only evaluate our model waveform on a coarser non-uniform grid, and then use the iterative evaluation described above in Sec.~\ref{sec:exp_eval} to evaluate the complex exponential of the phase. 

%As discussed above, the restrictions on the step size that we have derived from the phase are typically more restrictive than those on the relative error of the amplitude, with exceptions for large spins within a factor of two of the threshold value. 
By default we will use linear interpolation for the amplitude, with optional cubic spline interpolation. Both interpolation algorithms are currently using the open source \texttt{GSL} library \cite{gsl},
we do however expect a further speedup by replacing the \texttt{GSL} implementation by adding a standalone implementation of the required interpolations to our code.

We will now first discuss how to construct the non-uniform coarse frequency grid, and then the details of how to evaluate the waveform on the fine grid, first for spherical harmonics without mode mixing, and then for modes with mode mixing, which for the current \phXHM models concerns only the $\ell = 3, \vert m \vert = 2$ mode.

\subsubsection{Building the coarse frequency grid}

We assume that we are given an input frequency range $(f_{min},f_{max})$ where we need to evaluate the spherical harmonic modes of the waveform.  We wish to construct a non-uniform frequency grid, such that for every two successive frequency points the grid spacing $\Delta f(f)$ between them is sufficiently small to guarantee that the local phase error resulting from using linear interpolation between the coarse frequency points is smaller than a given threshold value $R$. We can then use eqs.~(\ref{eq:df_insp}, \ref{eq:df_mrd}) to compute $\Delta f$ as a function of the threshold $R$, the frequency $f$, the intrinsic parameters $(q, \chi_1, \chi_2)$, and the spherical harmonic mode $(\ell,m)$ under consideration.
The coarse grid will also depend on the desired grid spacing $df$ for the final 
uniform grid, since we build the coarse frequency grid such that the coarse points also belong to the fine grid. This simplifies the interpolation procedure for the complex exponential.

%Another very important input parameter is what we call the \textit{multibanding threshold} $R$, this is a measure of the error committed during the interpolation (see Sec.~\ref{sec:interp}) and allow us to control how restictive
%is the algorithm. 
%
Lower values for the threshold $R$ result in smaller errors, but higher computational cost. In Sec.~\ref{sec:results} we will 
compare different threshold settings and evaluate the computational cost
and compare the actual errors with the chosen threshold $R$.

As mentioned above in Sec.~\ref{sec:interp} we split the frequency range into three regions corresponding to the inspiral, merger and ringdown. For the practical implementation, instead of using the continuously varying $\Delta f(f)$ of expressions (\ref{eq:df_insp}, \ref{eq:df_mrd}), we work with a series of frequency bins where $\Delta f$ is fixed in each bin. The merger and ringdown parts have a much smaller dynamic range for $\phi^{\prime}(f)$ than the inspiral part (the phase ``flattens out'' from inspiral toward merger), and we just use one frequency bin for each region. Their spacings $\Delta f_{\mathrm{merger}}$ and $\Delta f_{\mathrm{RD}}$ are given by eqs.~(\ref{eq:dfmerger}) and (\ref{eq:dfringdown}-\ref{eq:df_amp_ringdown}) respectively. 

However, the inspiral part has a large dynamic range, and $\Delta f$ given by (\ref{eq:df_insp}) changes with a power law of $f^{11/6}$ so it also changes fast. The spacing that would accurately describe the whole inspiral part would be $\Delta f(f_{\mathrm{min}})$, however if we used this spacing for the whole region, we would be using many more points than what are really needed since $\Delta f$ increases so much for frequencies above $f_{\mathrm{min}}$. Therefore we use a varying number of frequency bins $N$, and we build each of them with a spacing $\Delta f_i$ twice larger than the previous bin. For the first bin we set $\Delta f_0 = \Delta f(f_{\mathrm{min}})$, thus
\begin{align}
\label{eq:df_firstbin}
\Delta f_0 &= \sqrt{\frac{2R}{c_{insp}}} f_{min}^{11/6}\\
\Delta f_{i} &= 2^i \:\Delta f_0, \hspace{1cm} i=1,2,...,N.
\end{align}
In practice we require that between two coarse points there is an integer number of fine frequency points, in consequence we modify $\Delta f_0$ such that
\begin{equation}
    \Delta f_0^\prime = \texttt{int}\left[\frac{\Delta f_0}{df}\right] \:df.
\end{equation}

Now that we have computed the spacing of each frequency bin, we need to compute the final frequency of each bin $f_{i,\mathrm{end}}$, which is the frequency that doubles the spacing $\Delta f$ of the current bin, i.e. we have to solve the equation $\Delta f (f_{i,\mathrm{end}}) = 2 \:\Delta f_{i}$. Inserting this into eq.~(\ref{eq:df_insp}) we obtain
\begin{align}
\label{eq:freq_factor}
    f_{i,\mathrm{end}}^{\frac{11}{6}} &= 2 \:f_{i,\mathrm{start}}^{\frac{11}{6}}, \\
    \frac{f_{i,\mathrm{end}}}{f_{i,\mathrm{start}}} &= 2^{\frac{6}{11}}.
\end{align}
We require that in a frequency bin there must be an integer number of coarse frequency points, and so we modify the end frequencies of each bin to 
\begin{equation}
    f_{i,\mathrm{end}}^{\prime} = \texttt{int} \left[  \frac{ f_{i,\mathrm{end}} - f_{i,\mathrm{start}}}{\Delta f_{i}} \right] \Delta f_{i}.
\end{equation}

With the above frequency factor we can estimate the number $N$ of bins that will be used in the inspiral. Since the inspiral regions ends at $f_{\mathrm{insp}}$, $N$ has to satisfy the relation
\begin{equation}
    \frac{f_{\mathrm{insp}}}{f_{\mathrm{min}}} =  \left(2^{\frac{6}{11}}\right)^N,
\end{equation}
and therefore we obtain
\begin{equation}
\label{eq:number_bins}
    N = \log_{2^{\frac{11}{6}}} \left( \frac{f_{\mathrm{insp}}}{f_{\mathrm{min}}} \right).
\end{equation}
Since $N$ is however the number of constant frequency bins for the inspiral, it has to be an integer, and
%
%In practice the expresions \ref{eq:df_firstbin}, \ref{eq:freq_factor}, \ref{eq:number_bins} above are not extrictly hold in order to simplify the interpolating algorithm. In first place, 
%
%
we modify $f_{\mathrm{insp}}$ such that 
\begin{equation}
    N = \log_{2^{\frac{11}{6}}} \left( \frac{f_{\mathrm{insp}}^{\prime}}{f_{\mathrm{min}}} \right) = \texttt{int}\left[{\log_{2^{\frac{11}{6}}} \left( \frac{f_{\mathrm{insp}}}{f_{\mathrm{min}}} \right)}\right].
\end{equation}

For the merger and ringdown regions, we proceed analogously to the inspiral region, and ensure that an integer number of fine grid points aligns with the coarse grid.
%
%Similar modifications are carried out for the merger and ringdown bin. 
Since this algorithm depends on the input values for $f_{min}$, $f_{max}$ and $df$, we perform several sanity checks to ensure that there is not any overlapping between regions. For example, if $f_{\mathrm{insp}}>f_{\mathrm{Lorentzian}}$ we skip the merger bin or if $f_{\mathrm{Lorentzian}}>f_{max}$ we skip the ringdown bin. 

\begin{figure}[ht]
    \centering
    \includegraphics[width=\columnwidth]{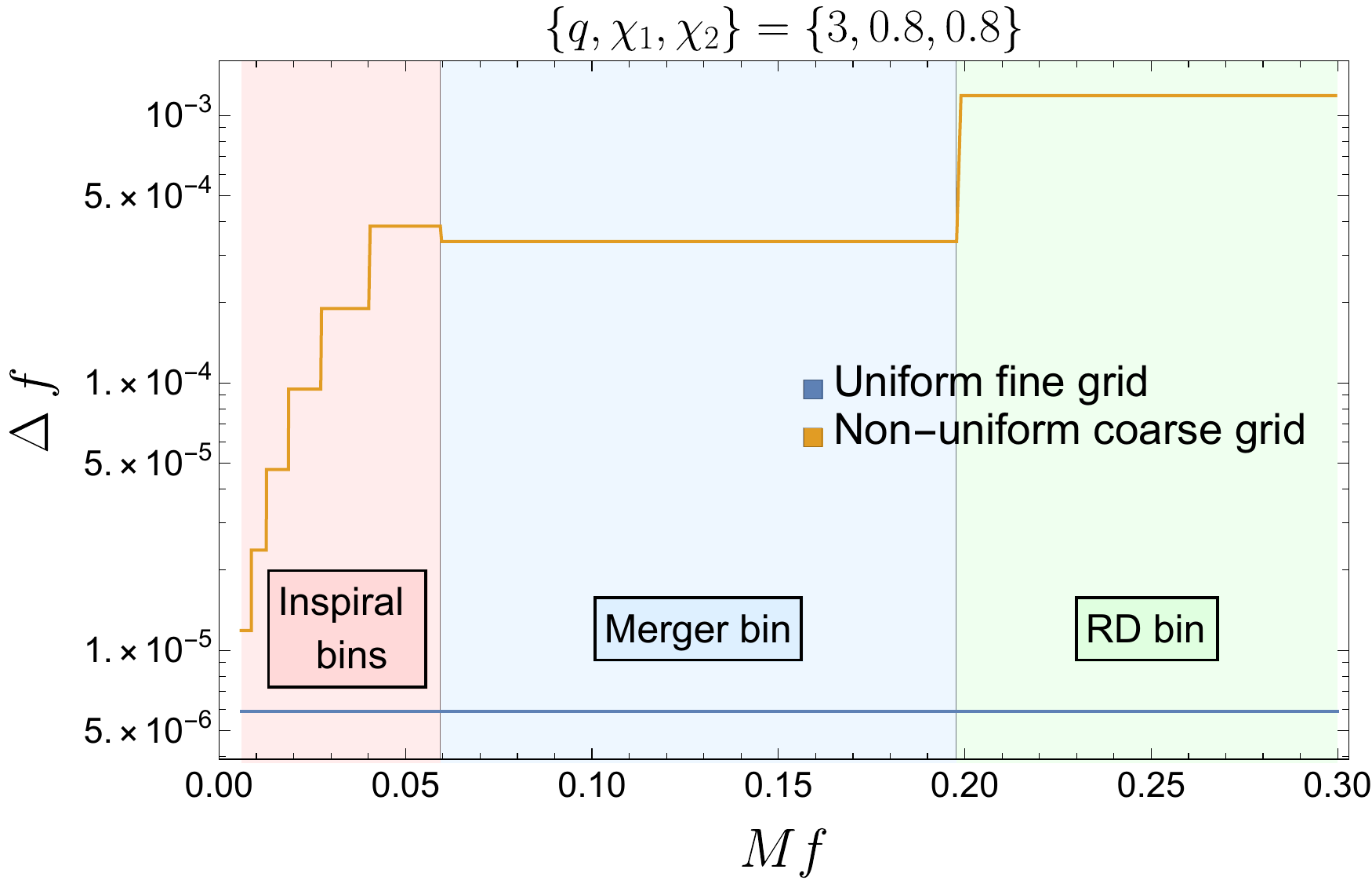}
    \includegraphics[width=\columnwidth]{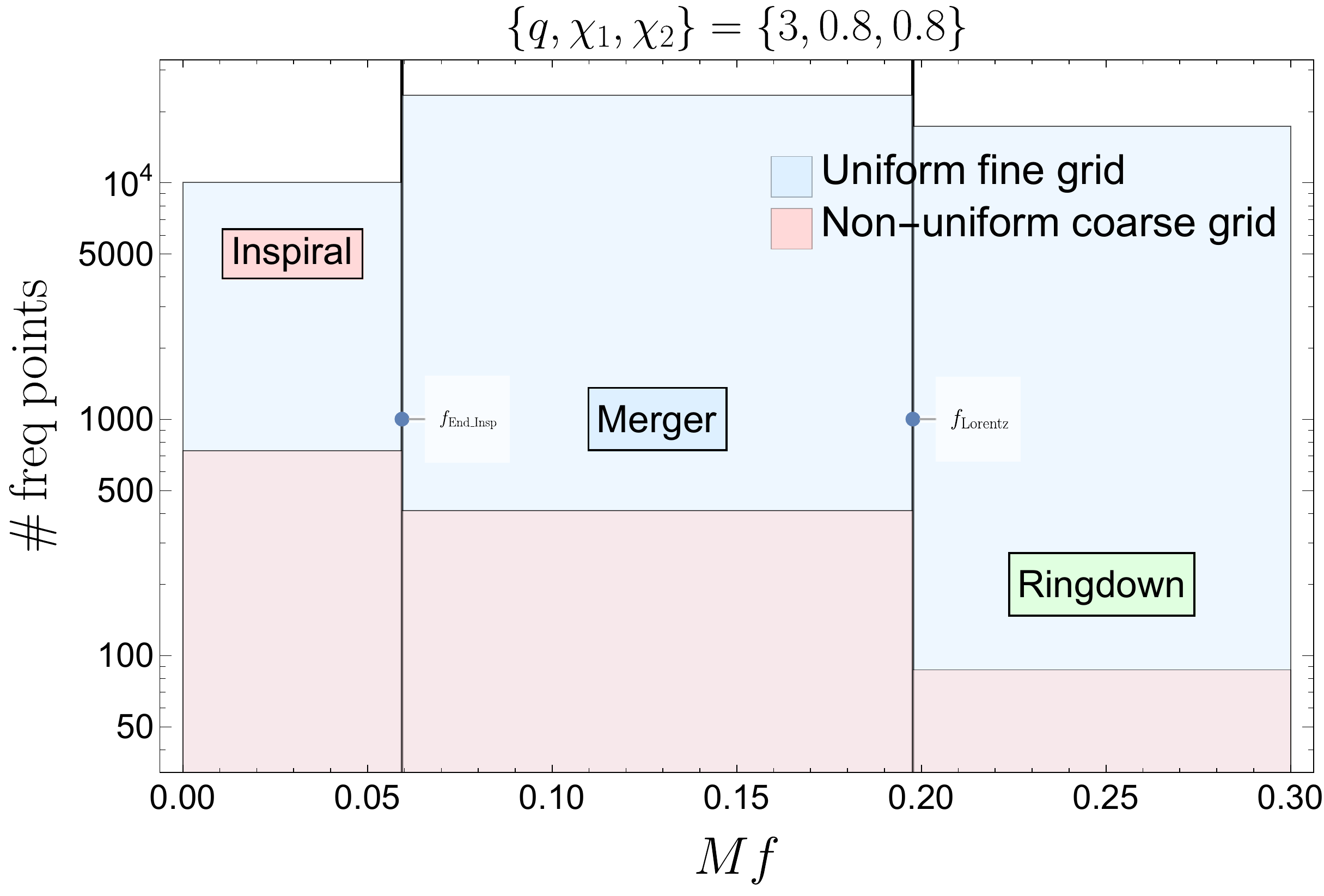}
    \caption{Comparison between uniform fine frequency grid and non-uniform coarse grid for the 33 mode for a high spin case. Top panel: The frequency spacing is shown for each frequency bin as a function of frequency. In the merger bin $\Delta f$ is smaller than the last one of the inspiral due to the need of resolving the dip of the Lorentzian. Bottom panel: The number of frequency points is shown for each frequency region. In a uniform grid most of the points lie in the merger and ringdown part where they are not so necessary, however this is corrected in the non-uniform grid.}
    \label{fig:coarse_grid}
\end{figure}

In Fig.~\ref{fig:coarse_grid} we compare the final non-uniform coarse grid with the uniform grid. In the top panel we can see how the frequency spacing $\Delta f$ increases for subsequent bins that constitute the inspiral part. In the case shown the merger bin has a slightly lower $\Delta f$ than the last inspiral bin in order to resolve the Lorentzian feature of the phase derivative. For other cases where the Lorentzian is less pronounced the limiting factor will be the derivative at the end of the inspiral and then the merger will have exactly twice the spacing of the last inspiral bin. The ringdown bin is the one with a coarser $\Delta f$ since there the phase derivative is practically flat.
In the bottom panel we show the number of frequency points for the inspiral, merger and ringdown parts. The uniform grid has most of its frequency points in the merger-ringdown part, which leads to an excessive computational cost in these regions, where far fewer points are required to capute the flatter behaviour of the phase derivative (see Fig.~\ref{fig:derivative}). For the  non-uniform grid most of the points are located in the inspiral part, where high
resolution is needed to describe the phase derivative. 

Now that we have described the non-uniform coarse frequency grid, the next step is to evaluate the model in this grid and carry out the interpolation to the fine grid. In this next step however, different procedures need to set up follow for the modes with and without mixing, as we discuss below.

\subsubsection{Evaluate the modes on the fine grid, with and without mode-mixing}

For the modes without mixing (at the moment all modes except $\ell = 3, \vert m\vert = 2$), the waveform modes are evaluated on the fine grid as follows:
\begin{enumerate}
    \item First the amplitude and phase are evaluated separately for the coarse frequency grid, which yields two 1D arrays, one for the amplitude and one for the phase.
    
    \item The complex exponential $e^{i \phi}$ is computed on the coarse grid.
    
    \item The fine uniform frequency grid is constructed with spacing $df$. 
    
    \item The complex exponential is interpolated to the final uniform frequency grid following the procedure described in Sec.~\ref{sec:exp_eval}.
    
    \item The amplitude is interpolated to the fine grid by using linear (optionaly third) order interpolation (using the \texttt{GSL} library). 
    
    \item The complex waveform $\tilde{h}_{lm}$ is constructed by multiplying the arrays for the amplitude and the complex exponential on the fine grid.
\end{enumerate}

%\subsubsection{Evaluate modes with mode-mixing}

For the modes with mixing (in our present implementation
of \phXHM this is only the $\ell = 3, \vert m\vert = 2$ 
mode) our procedure is slightly different from the modes without mixing. To handle mode-mixing, in the ringdown region the model is built in terms of spheroidal harmonics
instead of spherical harmonics, to simplify the waveform and avoid sharp features in the phase derivative and in the amplitude,
as discussed in detail in \cite{phenXHM}. After building the model waveform in terms of spheroidal harmonics, it is  then rotated back to spherical harmonics and connected with the inspiral part, which is directly modelled in terms of spherical harmonics. 
Performing our interpolation in terms of the spherical harmonics as for the other modes would require significantly higher resolution and increase computational cost.
%
%and we thus interpolate the model to the fine grid in the spheroidal representation, and then transform the interpolated waveform to spherical harmonics.
%
%since the mixing feature appears in the merger-ringdown part and the standard algorithm uses less points in this region. 
%The mixing is a complicated feature so we would need many more points to accurately described it, however as described in \cite{phenXHM}, this behaviour is much more smoother if the ringdown part is expressed in the basis of spheroidal harmonics. 
We thus use the same strategy as we have employed to construct the original model, and apply our multibanding algorithm separately to the inspiral region expressed in spherical harmonics, and to the ringdown part expressed in spheroidal harmonics, and then transform the latter to spherical harmonics once the fine grid values have been computed.
Our detailed procedure is as follows:
\begin{enumerate}
    \item We split the coarse frequency array into the spherical 
    part,  where we will perform the model evaluation and multibanding in terms of the spherical harmonics,
    and the spheroidal part, where we transform from the spheroidal to the spherical representation in the ringdown region.
    
    The start frequencies of the ringdown region for the phase and amplitude,
    $f_{RD}^{\mathrm{phase}}, f_{RD}^{\mathrm{amp}}$,
    are given in eq.~(5.2) in \cite{phenXHM}. Note that  $f_{RD}^{\mathrm{phase}}<f_{RD}^{\mathrm{amp}}$.
    For our multibanding algorithm we split between the ``spherical'' and  ``spheroidal'' coarse grids, where the spherical and spheroidal amplitude and phase are computed. There is some overlap between the frequency ranges of both
    in the interval ($f_{RD}^{\mathrm{phase}}$, $f_{RD}^{\mathrm{amp}}$),
    since 
    the spherical array goes up to $f_{RD}^{\mathrm{amp}}$, but the spheroidal one starts at $f_{RD}^{\mathrm{phase}}$, see step (8) below.
    
    \item Evaluate the spherical amplitude and phase in the spherical coarse array and evaluate the spheroidal amplitude and phase in the spheroidal coarse array, we get therefore four one-dimensional arrays. 
    
    \item Compute the complex exponential for the two coarse arrays of phases.
    
    \item Build the uniform frequency grid with spacing $df$ and split into spherical and spheroidal parts as above. 
    
    \item Interpolate the two arrays of complex exponential in their respective regions using the iterative procedure described in \ref{sec:exp_eval}. 
    
    \item Interpolate the two arrays of amplitude in their respective regions using linear (optionally third) order splin interpolation using the \texttt{GSL} library \cite{gsl}. 
    
    \item We have thus obtained four arrays: spherical amplitude and complex exponential evaluated in the spherical fine grid, and spheroidal amplitude and complex exponential in the spheroidal fine grid.
    
    \item Finally we combine amplitude and phase with different procedures in three frequency ranges:
            \begin{itemize}
            \item $f_{min} \leq f <  f_{RD}^{\mathrm{phase}}$: We directly multiply spherical harmonic amplitude and complex exponential.
            \item $f_{RD}^{\mathrm{phase}} \leq f <  f_{RD}^{\mathrm{amp}}$: We rotate to spherical the spheroidal complex exponential term (which requires the spheroidal amplitude), and then multiply the resulting spherical complex exponential with the spherical amplitude.
            \item $f_{RD}^{\mathrm{amp}} < f \leq f_{max}$: We first multiply the spheroidal amplitude and complex exponential and then transform to the spherical basis.
        \end{itemize}
%    \item The rotations to spherical need the 22 waveform. If the 22 mode has not been computed previously we evaluate it for the interval $\left( f_{RD}^{\mathrm{phase}}, f_{max} \right)$, if it is already compute we recycle the waveform so we gain in speed. 
\end{enumerate}

%%%%%%%%%%%%%%%%%%%%%%%%%%%%%%%%%%%%%%%%%%%%%%%%%%%%%%%%
\section{Results} \label{sec:results} 

%%%%%%%%%%%%%%%%%%%%%% Benchmarking %%%%%%%%%%%%%%%%%%%%%
\subsection{Computational performance} \label{sec:benchmark}

\begin{figure}[ht]
    \centering
    \includegraphics[width=\columnwidth]{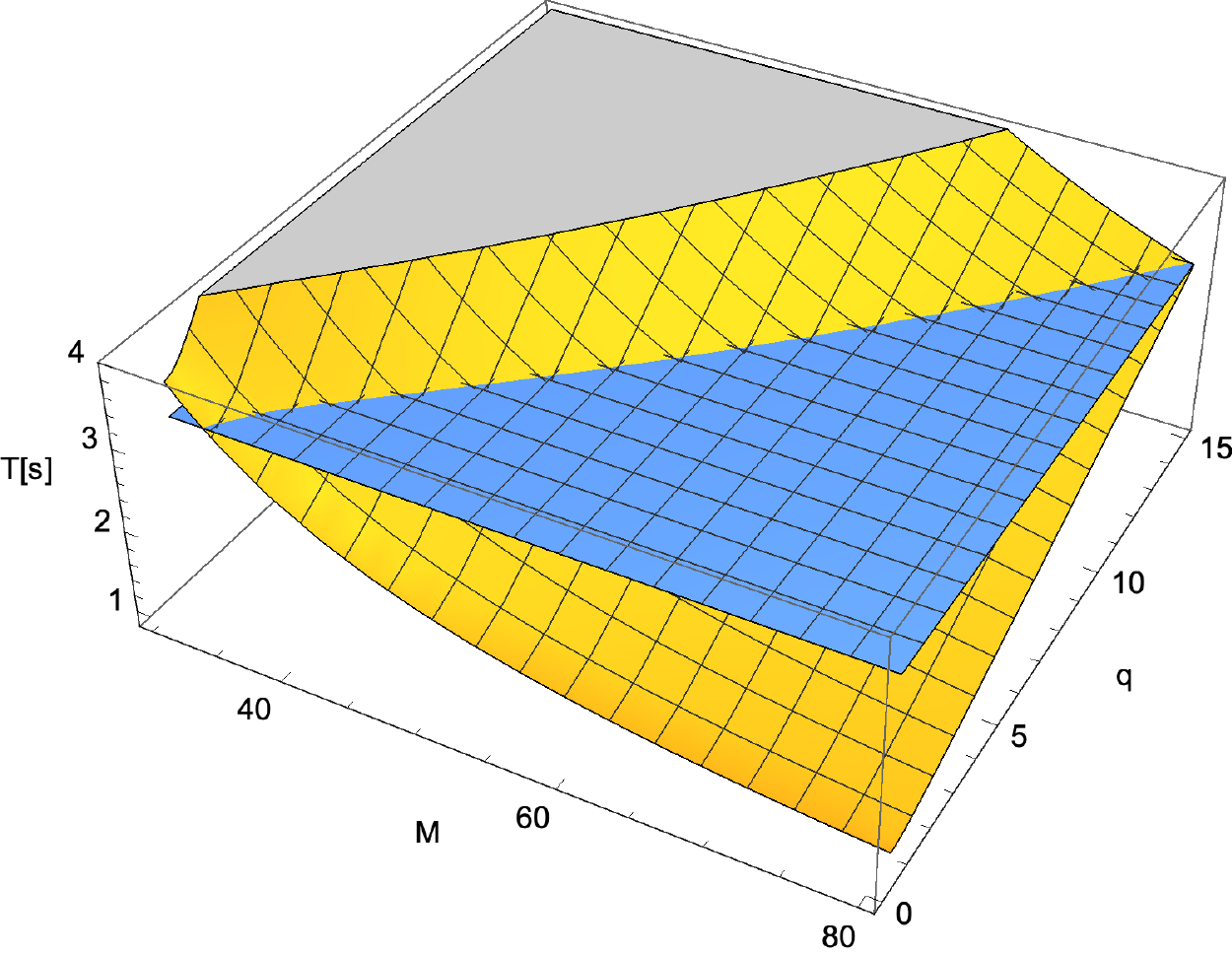}
    \includegraphics[width=\columnwidth]{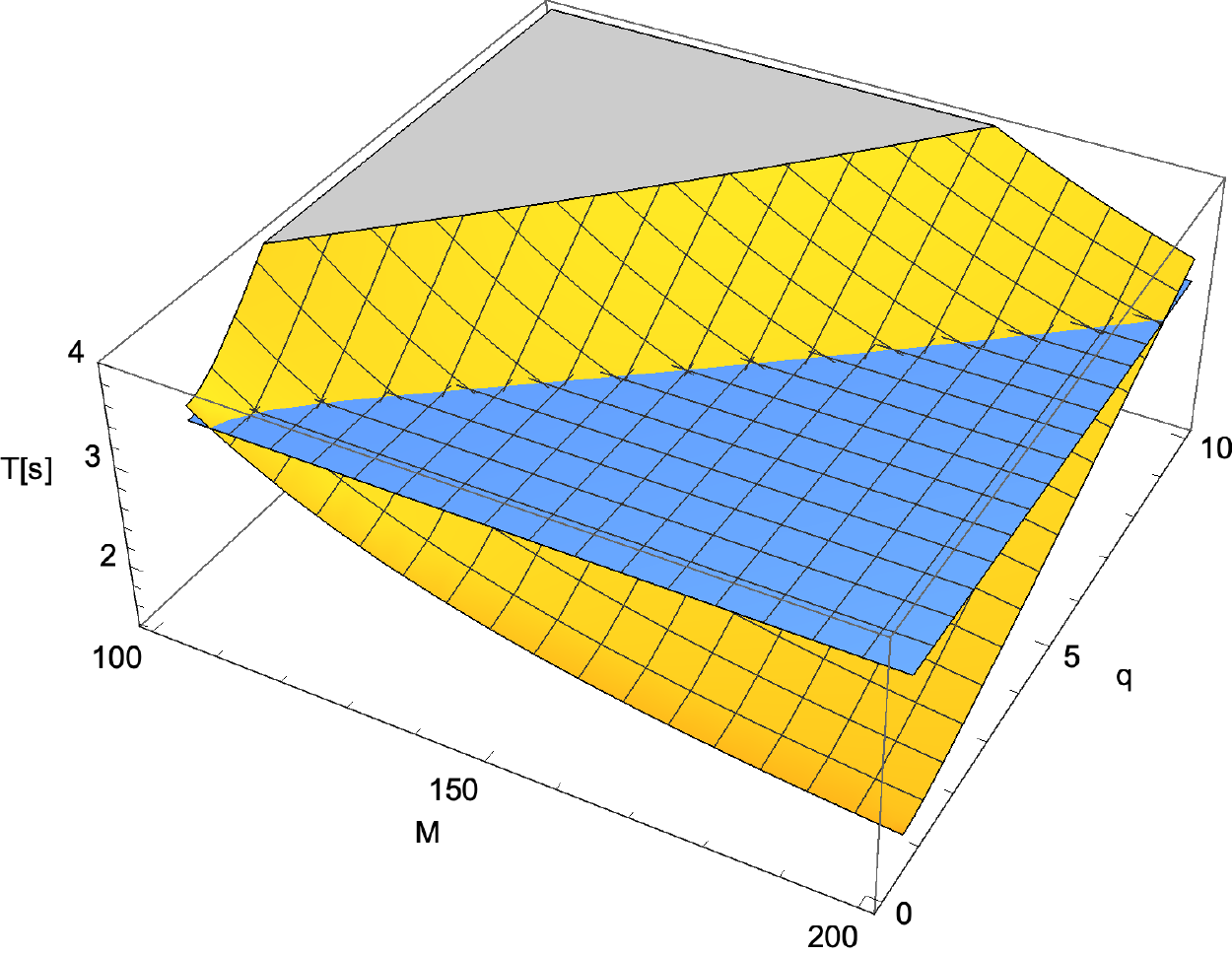}
    \caption{The approximate merger time observed by a detector with lower frequency cutoff at 20 Hz (upper panel) and 10 Hz (lower panel) is shown as a function of the total mass and the mass ratio of the system. The blue horizontal surface marks a duration of three seconds, which would allow for one second of buffer time between the signal duration and the length of the data segment. For a start frequency of 10 Hz only very high mass signals fit the time window.}
    \label{fig:T_vs_M_q}
\end{figure}

In first place we test the gain in speed due to multibanding and compare the results for different threshold values and for different spacings of the fine frequency grid. Note that the frequency spacing $df$ of the grid in the Fourier domain is related to the duration $T$ of the time segment that is analyzed by
\begin{equation}\label{eq:df_is_T_inv}
    df = \frac{1}{T},
\end{equation}
and thus longer signals require a smaller grid spacing.
To illustrate this dependency, in Fig.~\ref{fig:T_vs_M_q}
we show the approximate duration of a binary black hole coalescence signal as a function of mass and mass ratio.
To leading post-Newtonian order the duration in dimensionless units is given by
\begin{equation}
    T/M = \frac{5}{256 \eta  \, (\pi M f_0)^{8/3}},
\end{equation}
where $f_0$ is the frequency where the dominant spherical harmonic mode, $|\ell = \vert m \vert = 2$ enters the frequency band of the detector. Lower start frequencies thus imply much longer signals. 
In Fig.~\ref{fig:T_vs_M_q} we show results for two values of the lower frequency cutoff of the detector, $f_0 =$ 10 Hz, 20 Hz, the latter is what is typical for current compact binary parameter estimation, see e.g.~\cite{TheLIGOScientific:2016wfe,Chatziioannou:2019dsz}. The coalescence time is approximated with the TaylorT2 approximant at second post-Newtonian order spins aligned with the orbital angular momentum, and extreme Kerr values, adding
a time of $500 M$ in geometric units to account for merger and ringdown, in order to obtain an approximate upper limit on the duration. The figures focus on short signals, where time duration of 4 seconds is appropriate, and show the range of signals and templates in mass and mass ratio that fit into this time window.

\begin{figure}[ht]
    \centering
    \includegraphics[width=\columnwidth]{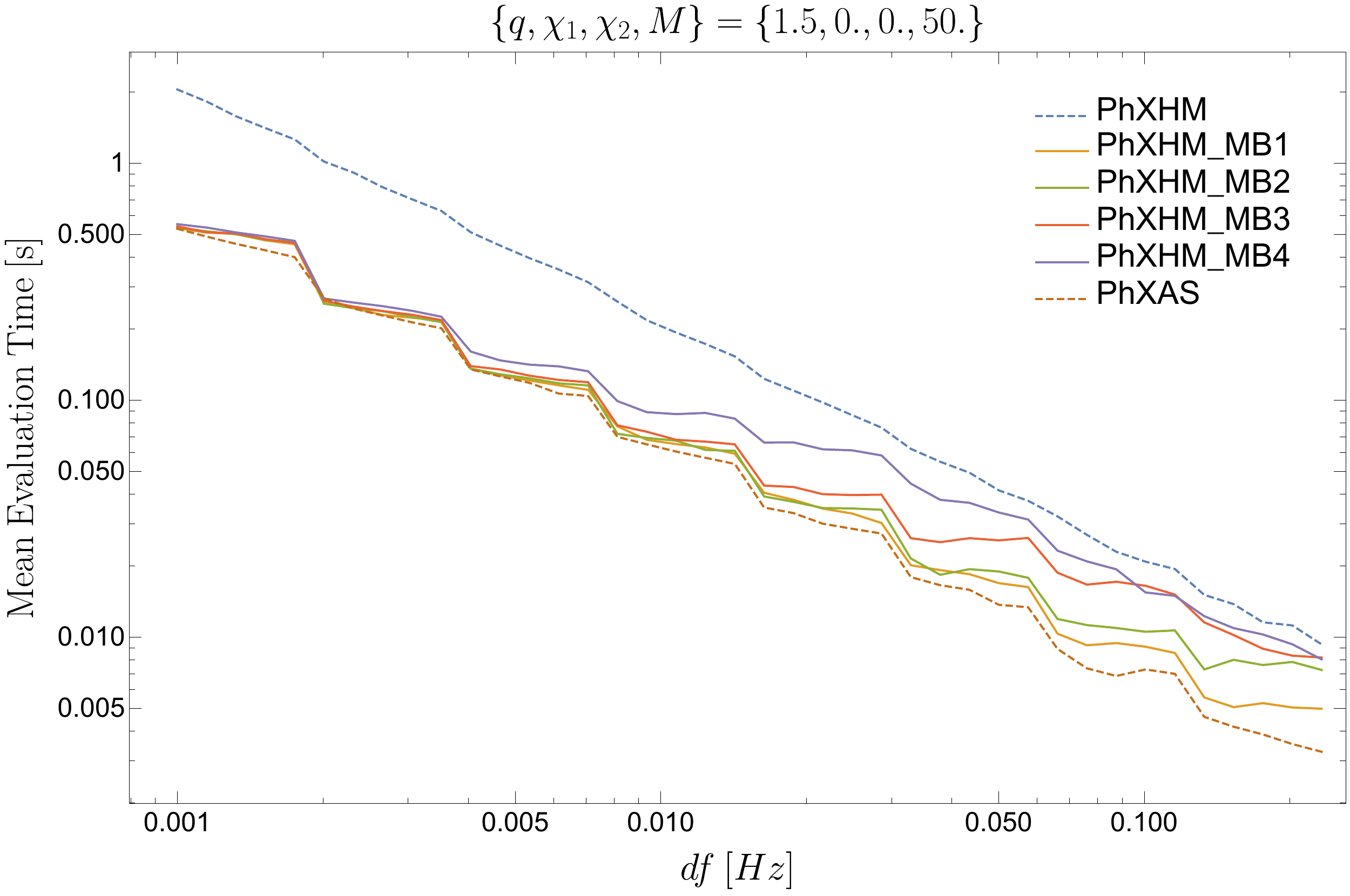}
    \caption{Evaluation time of the \texttt{LAL} code for waveforms with and without multibanding (solid and dashed lines respectively) for a total mass of 50 $M_{\odot}$. PhXAS denotes the quadrupole model, and the different ``PhXHM\_MB" items in the legend correspond to different values of the threshold $R$: $10^{-1}$, $10^{-2}$, $10^{-3}$ and $10^{-4}$. The evaluation time is averaged over 100 repetitions. The results are shown as a function of the spacing of the fine uniform frequency grid  $df$. They were obtained with the LIGO cluster CIT. }
    \label{fig:benchmark}
\end{figure}

We will now discuss an example case of a non-spinning system of black holes with total mass of $50$ solar masses and mass ratio $m_1/m_2 =1.5$, and evaluate the computational cost as a function of frequency spacing $df$.
In Fig.~\ref{fig:benchmark} we show the evaluation time of one waveform versus the spacing of the final uniform frequency grid. The frequency range spans from 10 to 4096 Hz and we fix the mass of the system to 50 $\mathrm{M}_{\odot}$. The dashed lines represent the waveforms generated without multibanding while the solid lines correspond to the multibanding version with  different values of the threshold: $10^{-1}$, $10^{-2}$, $10^{-3}$ and $10^{-4}$. 
First, we focus on the no-multibanding results, in principle we would expect that the higher modes model is 5 times slower than the 22-mode-only model because IMRPhenomXHM has 5 modes instead of just one. However it is a bit more expensive due to some particularities that are only present in the higher modes code, like the checks for the amplitude veto and mainly the extra steps needed to describe the mode-mixing of the 32 mode. 

Focusing now on the results with multibanding, notice that when $df$ is coarser the multibanding tends to equalize the no-multibanding. This is expected since for coarser $df$ we have less frequency points and then the coarse and fine grid tends to be similar and there is no gain by using the interpolation. Also it happens that the input $df$ may be larger than the $\Delta f$ of the coarse grid given by the multibanding criteria, in these cases we just evaluate in the frequency points of the fine grid and there is no gain in speed. In current LIGO-Virgo parameter estimation the highest $df$ that is used is 0.25 Hz, since $df$ is the inverse of the time duration of the signal as in eq.~(\ref{eq:df_is_T_inv}), and in practice the smallest duration considered, e.g. for high mass events with very short duration, is four seconds. Note however that as low frequency noise is reduced in detectors, and the lower cutoff frequency
for data analysis can be lowered, waveforms get longer 
and frequency spacings are reduced. 

On the contrary, when $df$ is very small we have a lot of points in the fine grid, then the interpolation is much more efficient and the multibanding has the highest gain in speed. 

The different values of the threshold behave as expected: larger values of the threshold are less accurate, but allow faster evaluation.
%
%the $0.1$ is the fastest one (although is also the lest accurate) and the $10^{-4}$ the slowest one (and most accurate). 
For small $df$ we observe however that the evaluation speed is almost independent of the threshold value. This is due to the fact that for small $df$ the evaluation of the model at the coarse grid points is computationally much cheaper than the subsequent interpolation to the fine grid points. Future optimization of our code will be required to address this issue and intend to reduce the computational cost of the interpolation to the fine grid.

%The explanation for this is quite technical and for the shake of clarity let's consider just two values of $df$ with $df_1 < df_2$ and two values of the threshold $0.1$ and $10^{-4}$. There are two operations that contribute to the evaluation time, one is the evaluation of the model in the coarse array and the other is the number of interpolations between two coarse points (determined by $df$). A very naive formula to get the evaluation time would be 
%\begin{align}
%    t_{\mathrm{eval}}(df, R) = &\mathrm{\:length \:coarse \:array}\\ +& %\mathrm{\:interpolations \:per \:coarse \:point}.
%\end{align}
%
%Here we can see that a decrease in the length of the coarse array could be compensated if the number of interpolations per coarse point increase a lot. In general the threshold $0.1$ is faster than $10^{-4}$ because the length of the coarse array is lower than for $10^{-4}$ and the number of interpolations for $df_1$ and $df_2$ is not very different. However, when $df_1$ is very small $df_1 \ll df_2$ the number of interpolations between coarse points increases a lot for $df_1$ respect to $df_2$ and compensate the gain in speed of the shorter coarse array. 

\begin{figure}[ht]
    \centering
    \includegraphics[width=\columnwidth]{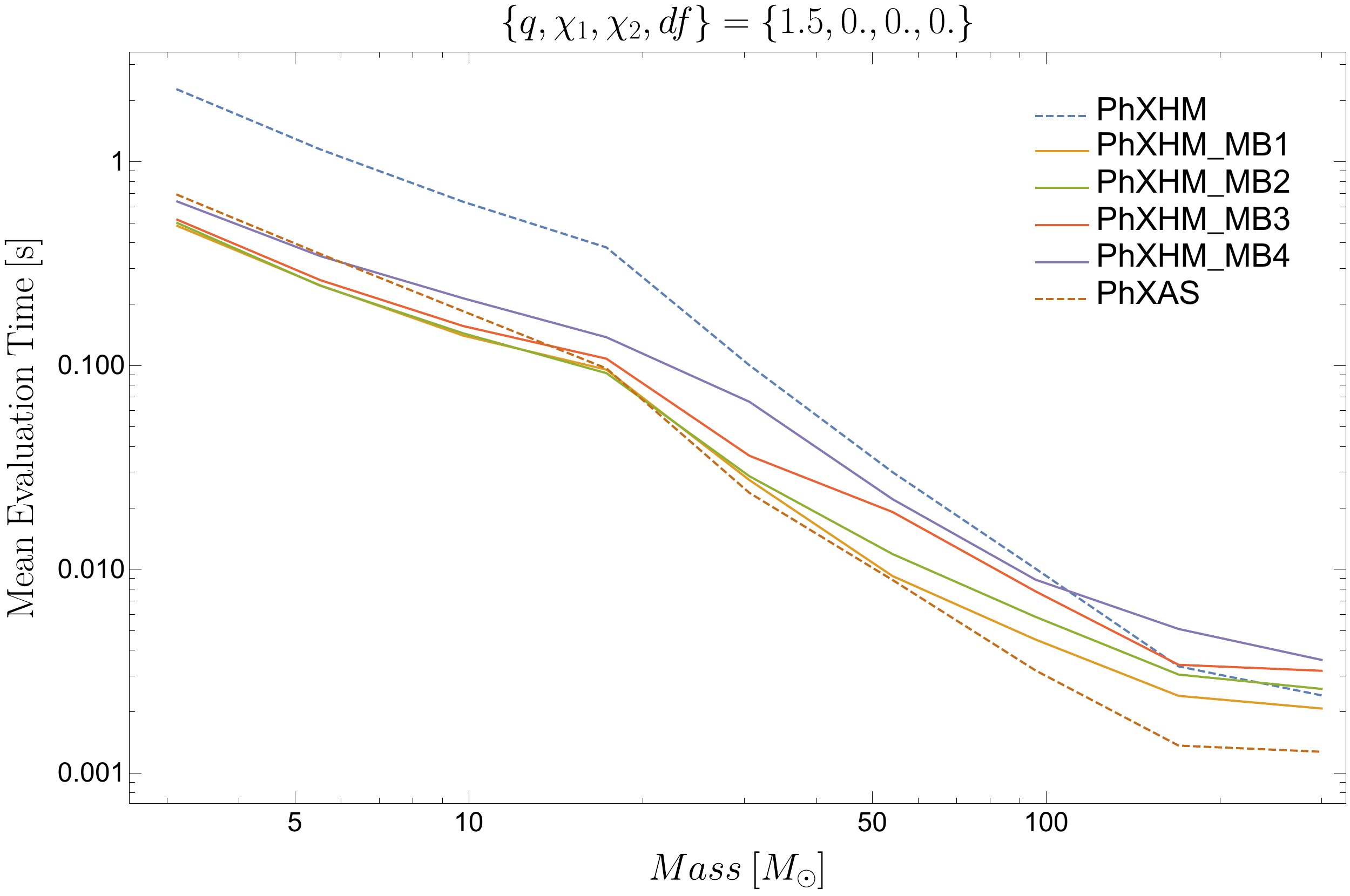}
    \caption{Evaluation time versus total mass of the system, choosing a fine grid spacing $df$ that corresponds approximately to the inverse merger time at this mass ratio, as is common on data analysis applications (we denote this behaviour by writing $df=0$). The evaluation time is averaged over 100 repetitions. Result were obtained with the LIGO cluster CIT.}
    \label{fig:benchmark_df0}
\end{figure}

We now show the dependence of the evaluation time on the total mass of the system. In this case the spacing of the fine grid $df$ is computed by the \texttt{LAL} function \texttt{XLALSimInspiralFD} which adapts the $df$ accordingly to an internal estimation of the time duration of the signal which depend on the lower cutoff $f_{min}$, chosen here as $f_{min}=10 Hz$, and the mass of the system. This is similar in spirit
to the estimate of the merger time that we have used in Fig.~\ref{fig:T_vs_M_q}. In Fig.~\ref{fig:benchmark_df0} we see qualitatively the same results than when simply scaling $df$ as in Fig.~\ref{fig:benchmark}. The multibanding is more efficient for lower masses where the duration of the signal is lower and therefore a smaller $df$ is used.

\subsection{Accuracy} \label{sec:accuracy}

\begin{figure*}[htpb]
    \centering
    \includegraphics[scale=0.3]{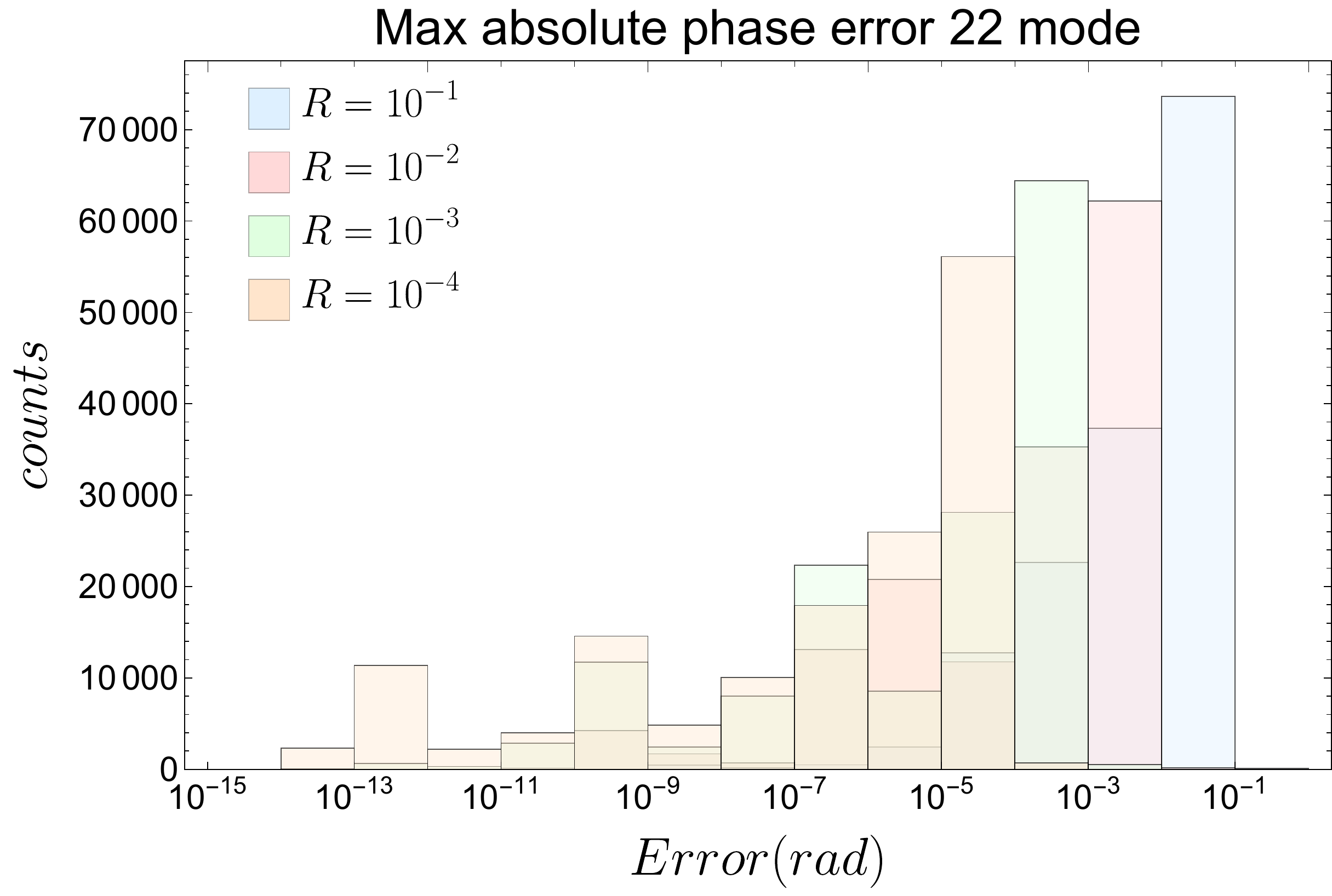}\includegraphics[scale=0.3]{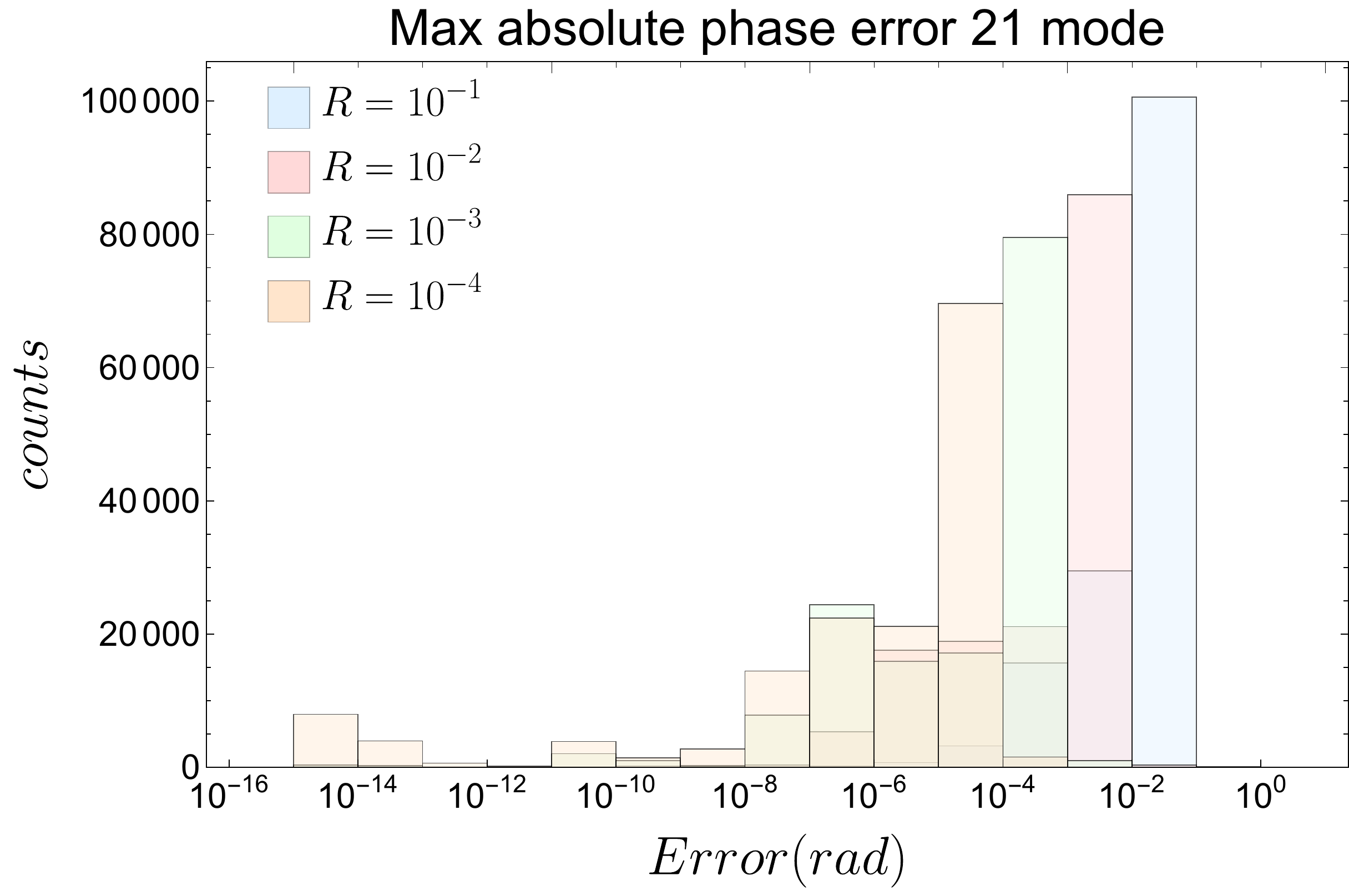}
    \includegraphics[scale=0.3]{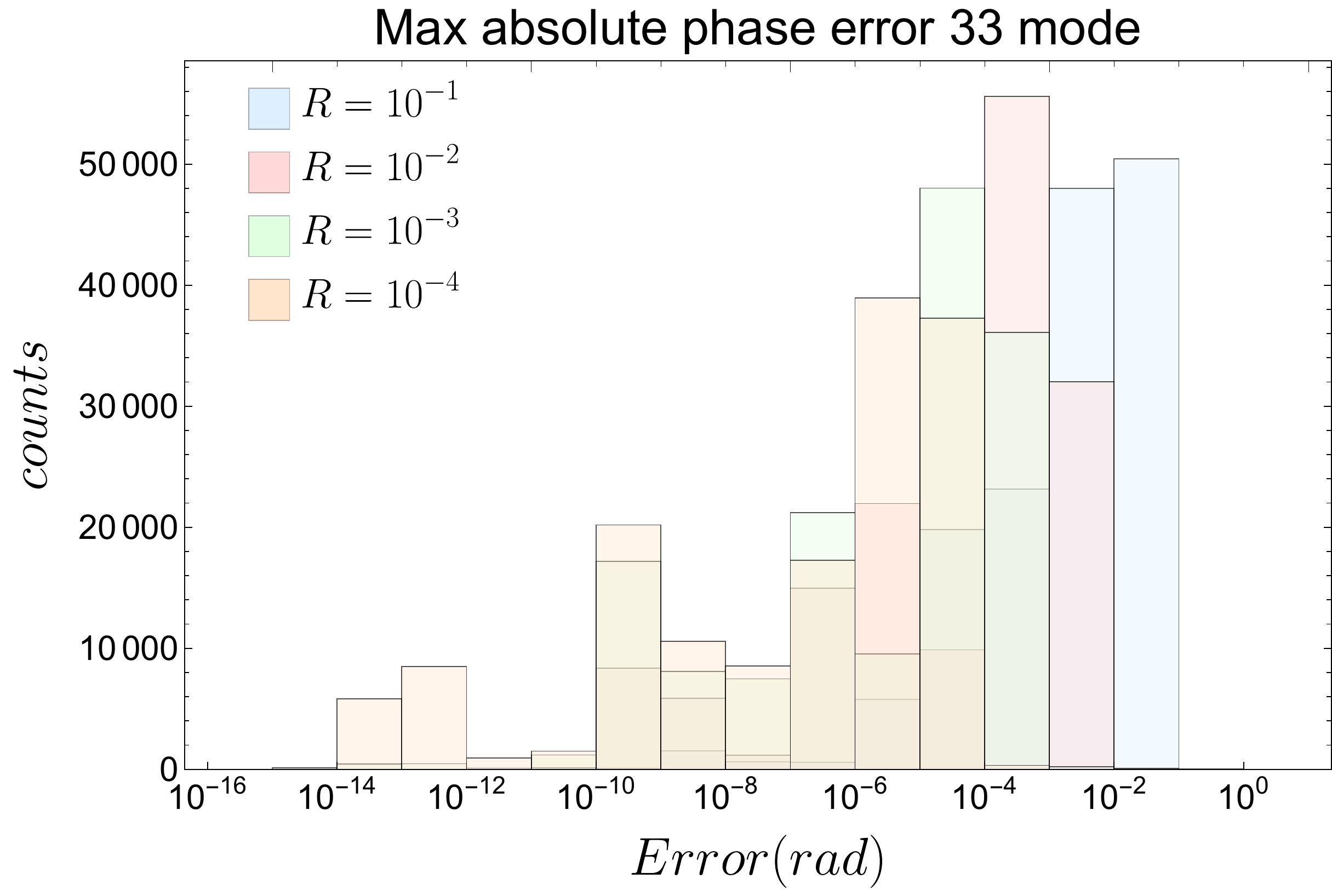}\includegraphics[scale=0.3]{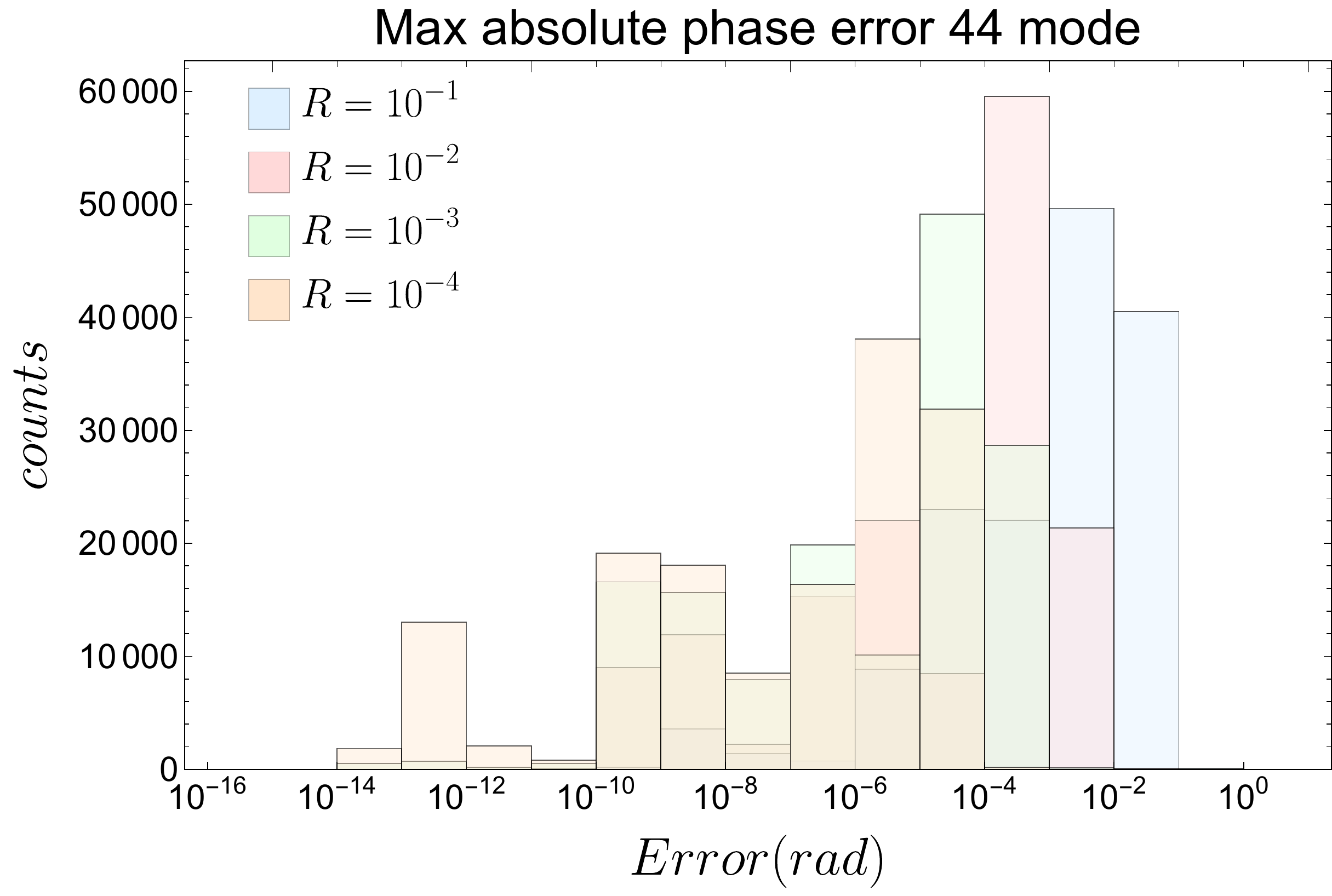}
\caption{Maximum absolute error for the phase of the $(l,m)$ mode between the multibanding and no-multibanding waveforms for four values of the threshold $R$. The threshold $R$ can be interpreted as an approximate upper limit for the maximum absolute error introduced in the phase.}
    \label{fig:error_phase}
\end{figure*}

\begin{figure*}[htpb]
    \centering
    \includegraphics[scale=0.3]{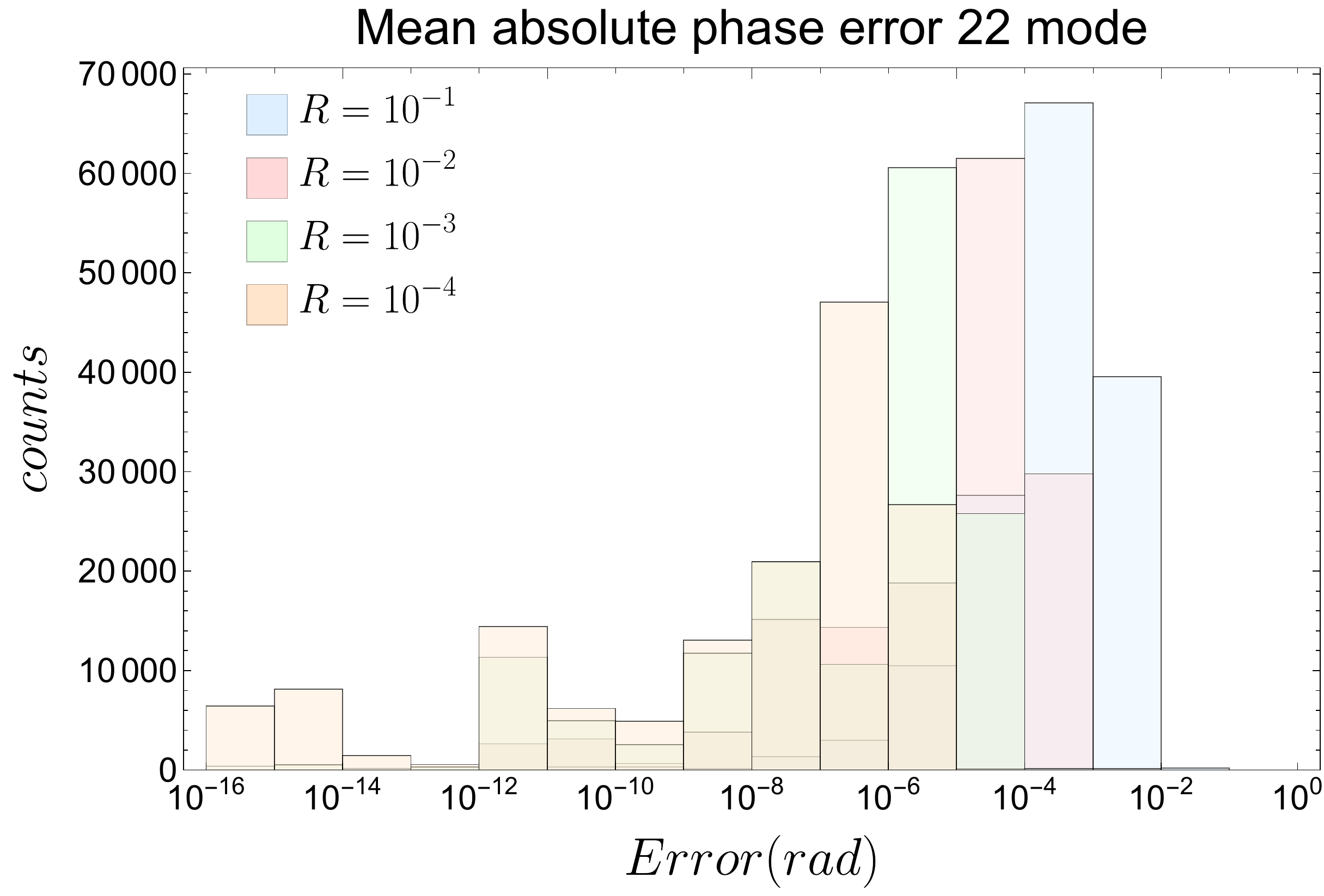}\includegraphics[scale=0.3]{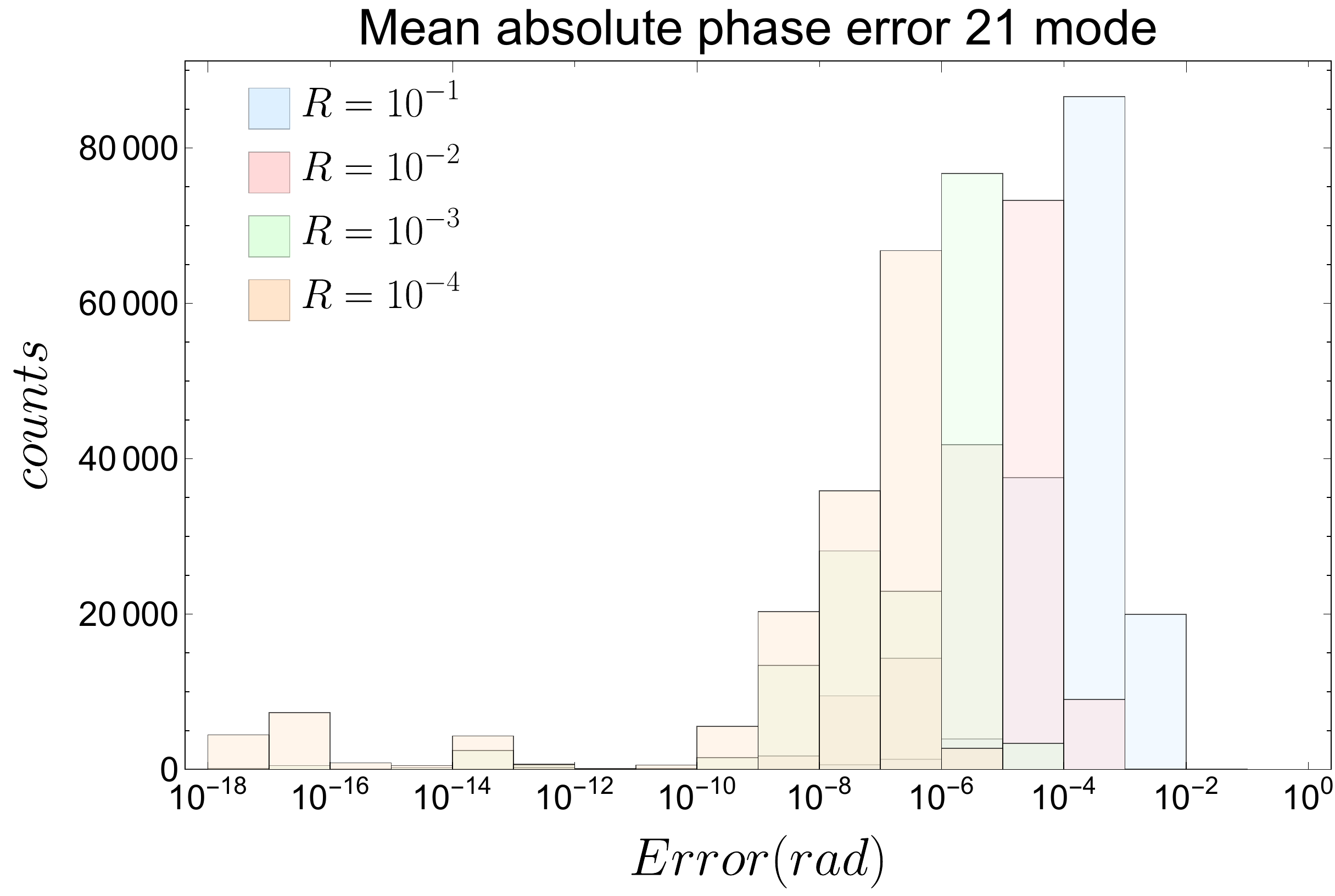}
    \includegraphics[scale=0.3]{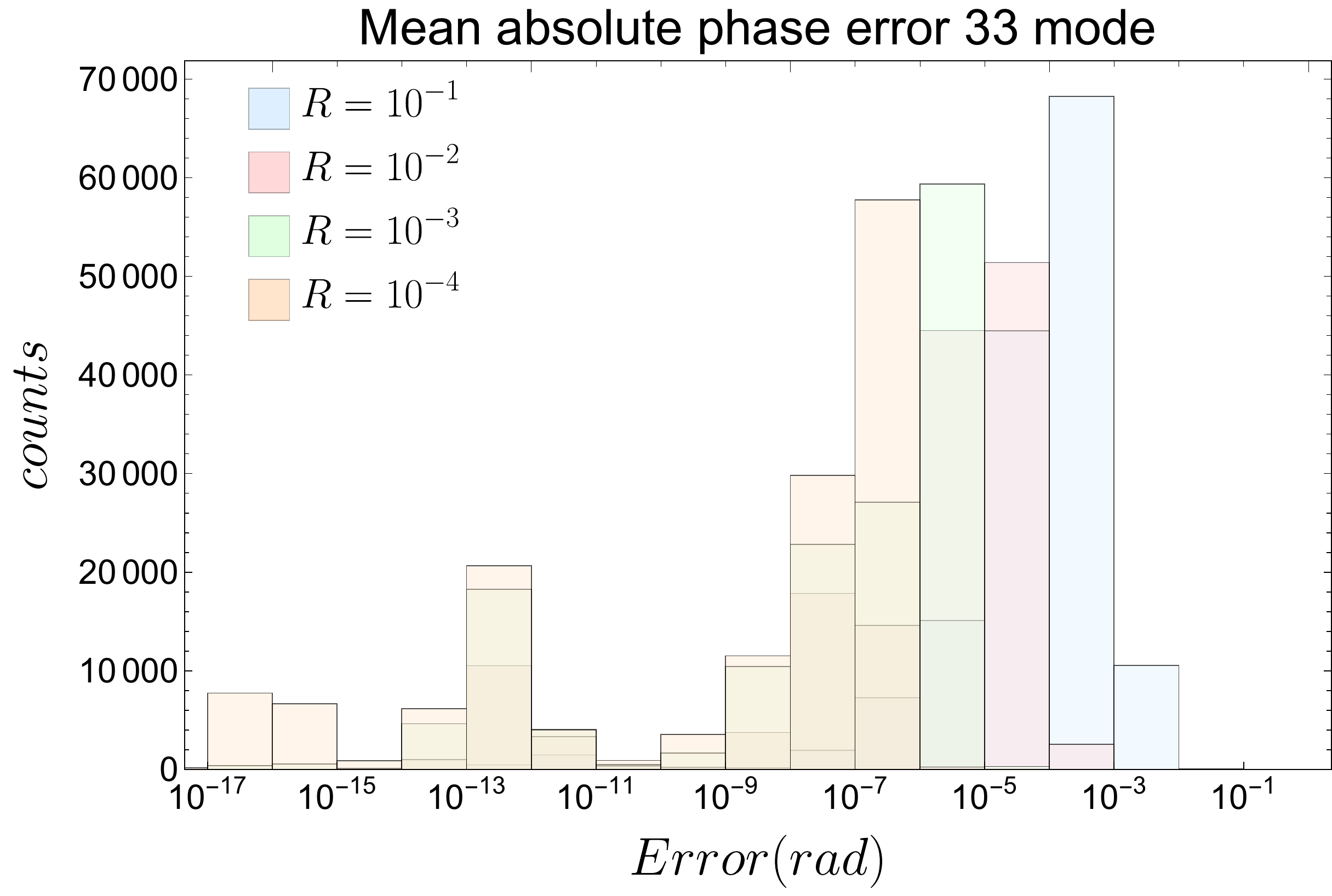}\includegraphics[scale=0.3]{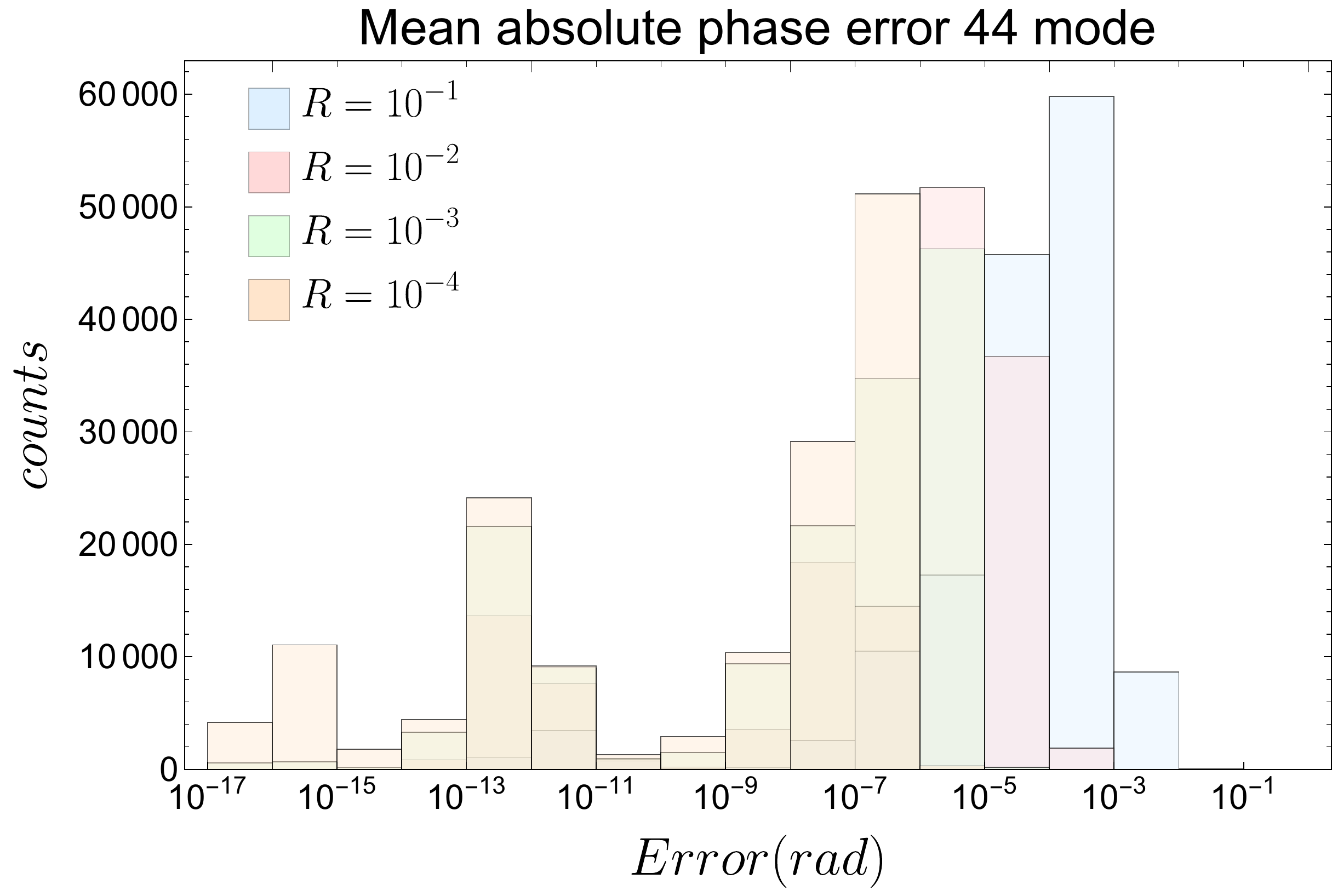}
\caption{Absolute error for the phase averaged over the frequency array for the $(l,m)$ modes between the multibanding and no-multibanding waveforms for four values of the threshold $R$. The threshold $R$ can be interpreted as an approximate upper limit for the absolute error introduced in the phase.}
    \label{fig:error_phase_mean}
\end{figure*}

\begin{figure}[ht]
    \centering
    \includegraphics[width=\columnwidth]{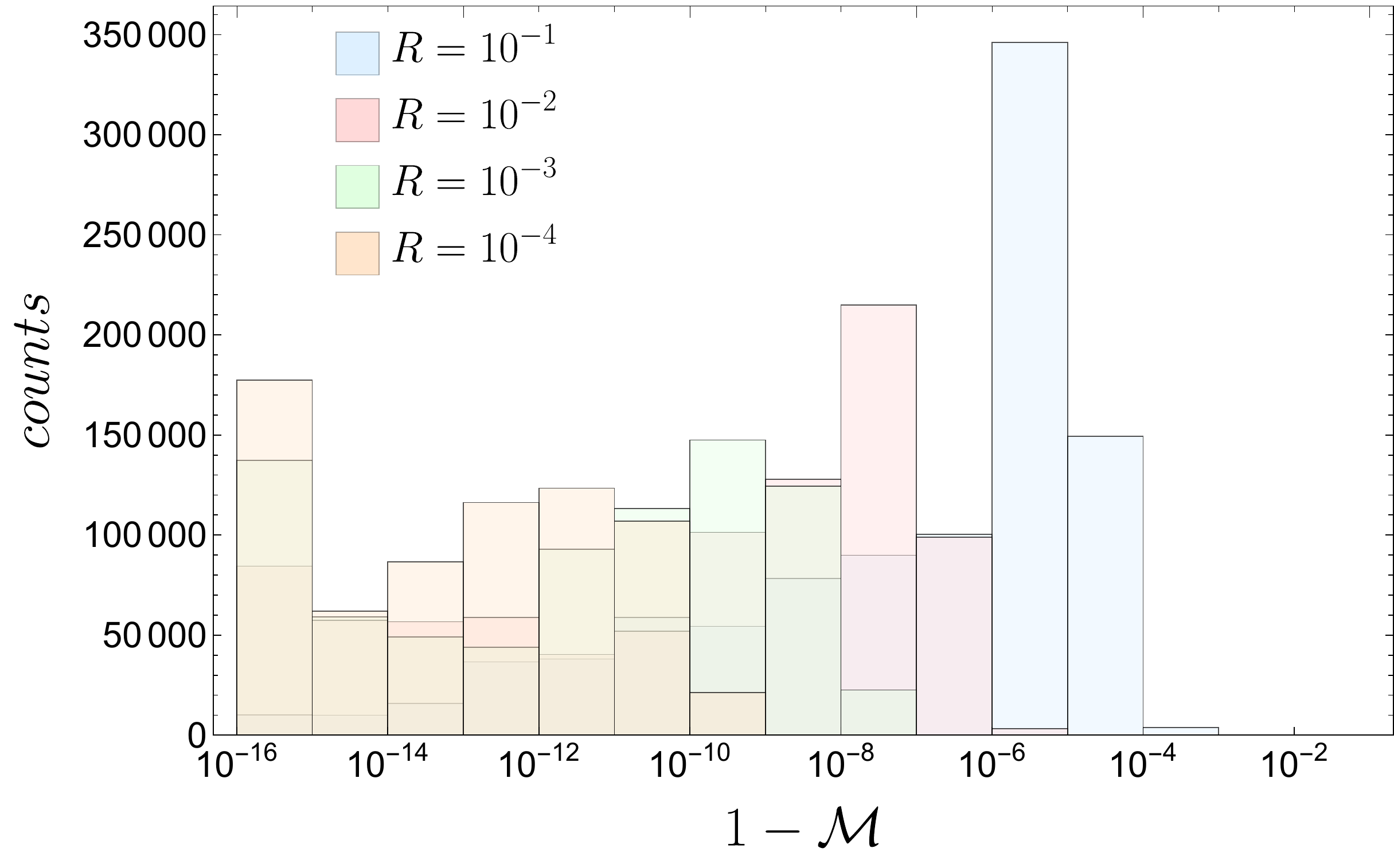}
\caption{Mismatches between the no-multibanding and multibanding waveforms for one million random configurations across parameter space. We show the results for four different values of threshold. The configurations are choosen randomly with $q \in [1, 1000]$, $\chi_{1,2} \in [-1, 1]$, $\mathrm{M_t} \in [1, 500] M_{\odot}$, $df \in [0.01, 0.3]$ Hz, $\iota \in[0,\pi]$, $f_{min}=10Hz$, $f_{max}=1024Hz$. The mismatch is computed for the $h_{+}$ polarization.}
    \label{fig:mband_matches}
\end{figure}

%%%%%%%%%%%%%%%% Error within thresholds %%%%%%%%%%%%%%%%%%%%%%
In this section we discuss the accuracy of the multibanding algorithm as well as compare different choices of the threshold and motivate the choice of the default value for the model. In section ~\ref{sec:interp} we explained that the non-uniform coarse grid is built such that the error in the phase (of a single mode) is below a threshold $R$. %Here we show the results for just one mode, see Appendix ~\ref{} for the rest of the modes as well as for the final multimode waveform. 
%We compute the waveform with and without multibanding for 141650 random configurations with $m_1 \in [1,100]$, $m_2 \in [1,100]$, $\chi_1 \in [-1,1]$, $chi_2 \in [-1,1]$, $df \in [0.01,1]Hz$ and $\texttt{phiRef} \in [0,2\pi]$. For the multibanding we compute again for four different threshold values. In Fig.~\ref{fig:error_phase}, we show the maximum absolute error for the whole uniform frequency array as well as the mean value for the array. We see that while most of the frequency points for all the configurations are indeed below their corresponding threshold, there are several cases where the maximum error is above. This discrepancy comes from the approximations that the algorithm does, like using the TaylorF2 phase to approximate the phase in the inspiral to compute $\Delta f$. This is known to not be very accurate for extreme spins and also for cases with high mass the inspiral starts at higher geometrical frequencies where the TaylorF2 is not so accurate. 
% \begin{figure}[ht]
%     \centering
%     \includegraphics[width=\columnwidth]{"Plots/Max absolute phase error".pdf}
%     \includegraphics[width=\columnwidth]{"Plots/Mean absolute phase error".pdf}
%     \caption{Top panel: Maximum absolute error for the phase of the lm mode between the multibanding and no-multibanding waveforms for four values of the threshold $R$ and 140000 configurations. Bottom panel: same than above but showing the mean absolute error across the frequency array instead of the maximum. }
%     \label{fig:error_phase}
% \end{figure}
To check this we compute the waveform with and without multibanding for 150000 random configurations in the parameter range $M_{tot} \in [1, 500] M_{\odot}$, $q\in [1,1000]$, $\chi_{1,2}\in [-1, 1]$, and $df  \in [0.01, 0.25]$Hz. For the multibanding we compare again four different threshold values. In Fig.~\ref{fig:error_phase}, we first show the maximum absolute error for the whole uniform frequency array for all the modes except $\ell=3, m=2$, 
where mode mixing needs to be taken into account for interpreting results as discussed below. We see that for most cases the maximum error is indeed below the threshold. However there can be special configurations where a few frequency points may give an error above the threshold. These few cases correspond typically to configurations where the approximations employed by the algorithm are less accurate, e.g.~using the TaylorF2 phase to approximate the phase in the inspiral to compute $\Delta f$ for extreme spins or for cases with high mass, where the inspiral starts at high frequencies where the TaylorF2 is again less accurate. In Fig.~\ref{fig:error_phase_mean} we show the mean error, averaged over the frequency array, always remains below the thresholds. We will also compute mismatches between the original and the interpolated below, and find that we indeed achieve acceptably low values of the mismatch, see Fig.~\ref{fig:mband_matches}.

%Another reason for artefacts is simply a shortcoming of our test: 
%While for data analysis applications the full waveform is used, we
% need to reconstruct phase to compute the phase error. We take the phase by computing the argument of a complex frequency series and do the unwrapping such that the phase monotonically increases, but sometimes the unwrap does not works properly if the frequency spacing is too large. We also noticed that this method has difficulties when there is big drop in the amplitude (which affect mainly to the 32 mode due to the mode-mixing).
% We observe that the 32 mode is the one with larger errors, and the reason is that in the ringdown we are computing the phase in the basis of spherical harmonics, but the multibanding algorithm is applied to the phase in spheroidal harmonics and therefore they should not be directly compared. If instead of looking at the maximum phase error we look at the average error over the frequency array the result are much nicer and we consider those more informative of the overall accuracy of the algorithm, see~Fig.\ref{fig:error_phase_mean}. 
% In spite of we can not claim that the phase error will be always below the corresponding threshold we still consider that these parameter is a very good indication of the accuracy of the algorithm. In the following we will study the effect of the errors in the mismatch, which truly determine the accuracy of the waveform. 

\begin{figure}[ht]
    \centering
    \includegraphics[width=\columnwidth]{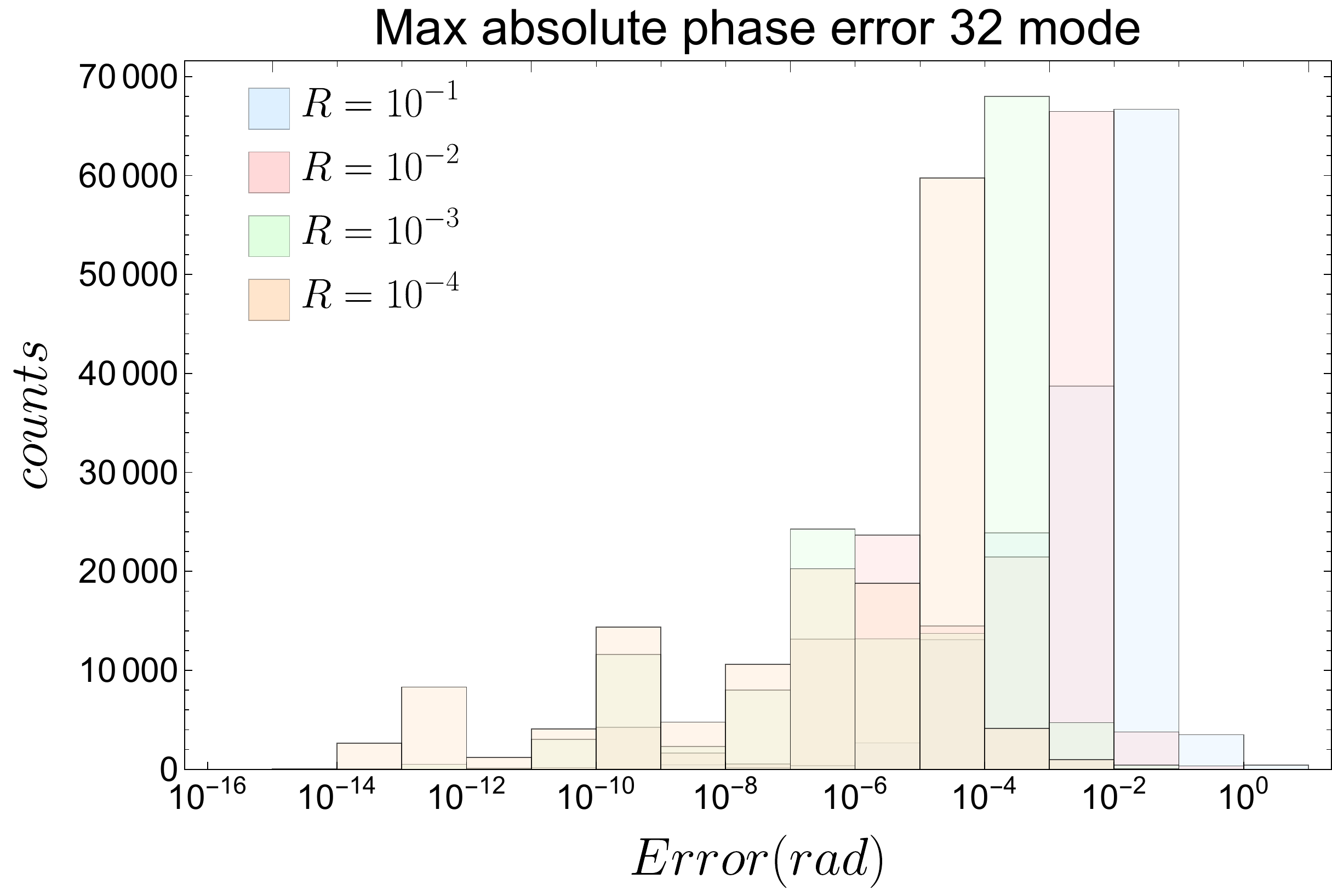}    \includegraphics[width=\columnwidth]{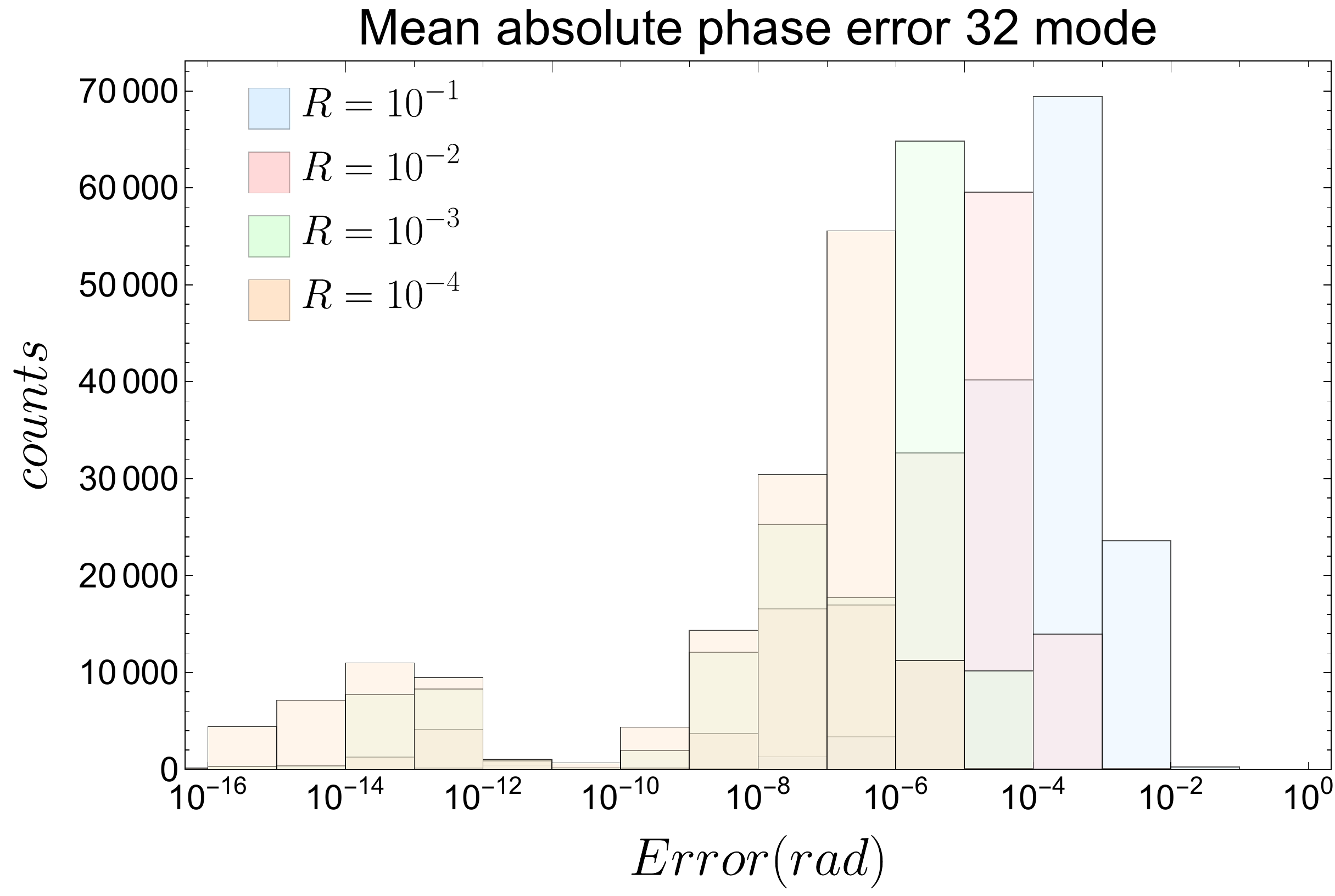}
    \caption{Maximum absolute error (top) and averaged over the frequency array (bottom) for the phase of the $(3,2)$ mode between the multibanding and no-multibanding waveforms for four values of the threshold $R$.}
    \label{fig:error_phase_32}
\end{figure}

Now we consider the 32 mode, where mode mixing with the 22 mode is present. In
 Fig.~\ref{fig:error_phase_32} we show the results for the same test as shown in Fig.~\ref{fig:error_phase} for the other modes.
The interpretation of the results is however different now, since
the ringdown of the 32 mode, where mode mixing is present, is modelled 
and interpolated in terms of the spheroidal harmonics, see \cite{phenXHM},
where the waveform phenomenology is much simpler than in terms of spherical harmonics.

%The reason is because for the ringdown part the multibanding algorithm is applied to the waveform in the basis of spheroidal harmonics where it is more smoothly behaved and therefore the threshold value is only valid for the phase in spheroidal. However here we are substracting the spherical phase that of course is not required to satisfy the criteria of the threshold. This is particularly true when a sharp feature is present in the waveform. That is why only for the 32 mode we see an "excessive" number of cases above the threshold. 

\begin{figure}[ht]
    \centering
    \includegraphics[width=\columnwidth]{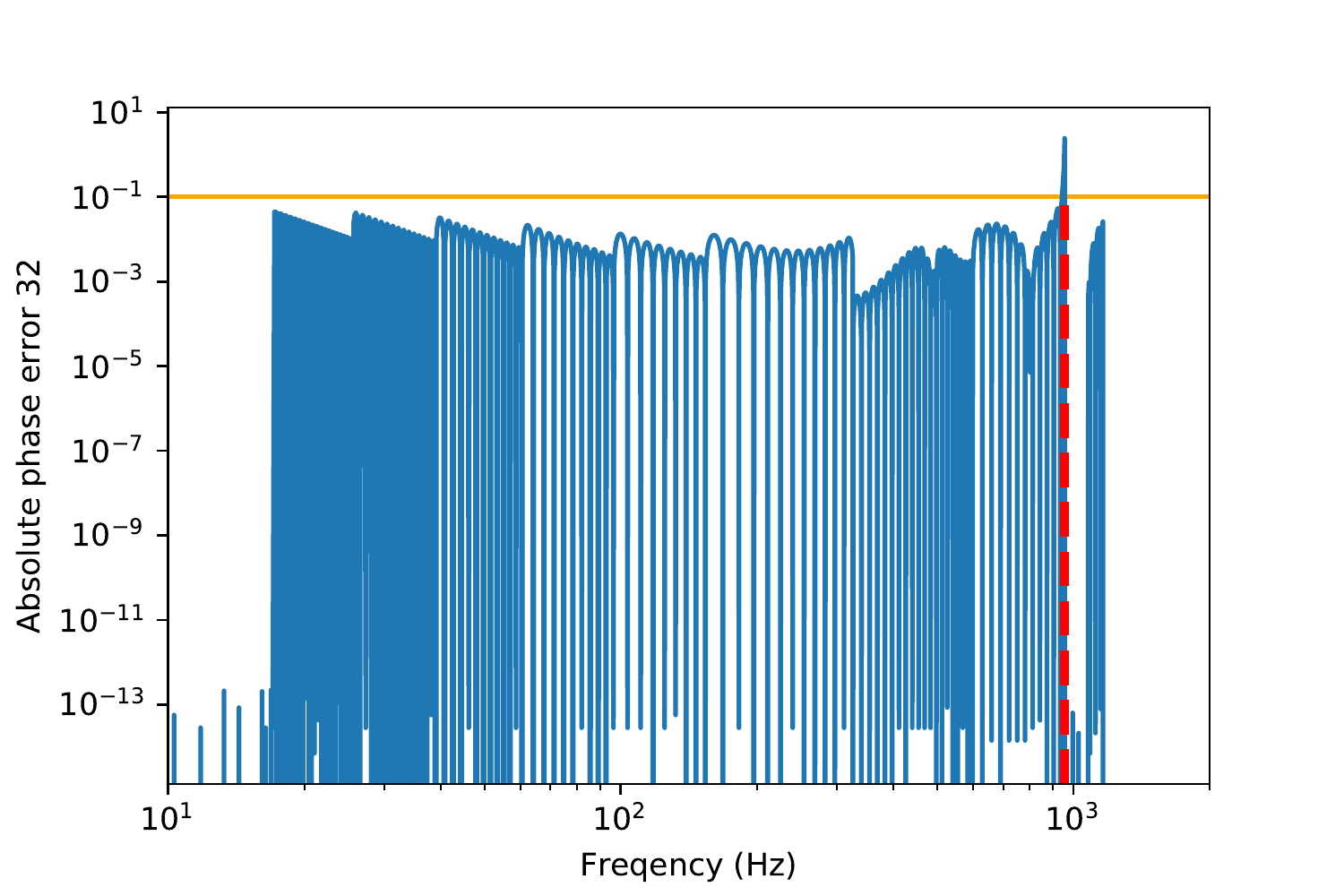}    \includegraphics[width=\columnwidth]{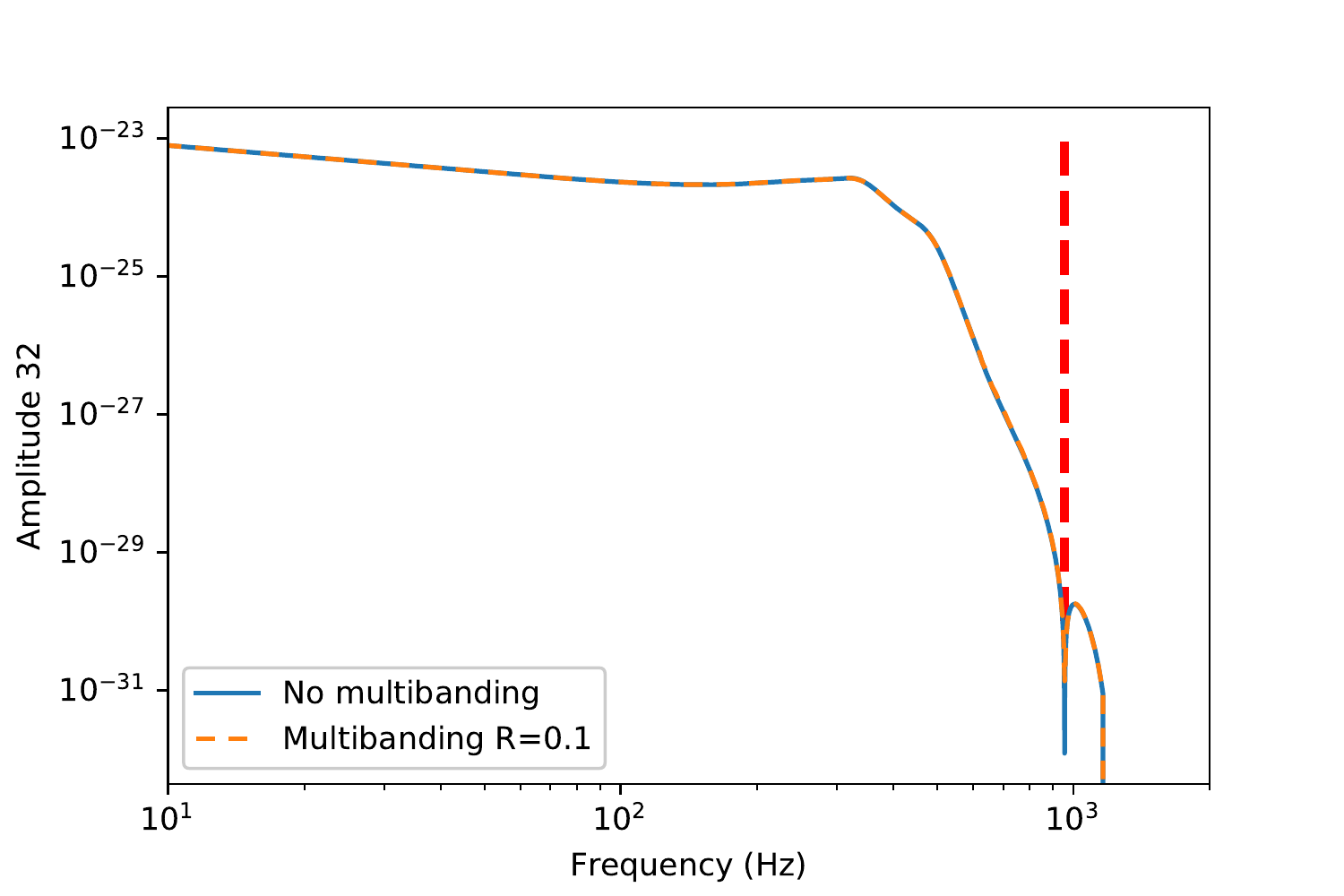}
    \caption{Example case that produces a maximum phase error above the threshold for the 32 mode. The parameters for this case are $m_1=35.7 M_{\odot}$, $m_2=16.6M_{\odot}$, $\chi_1=0.33$, $\chi_2=-0.54$. Top panel: Absolute phase error between the no-multibanding and multibanding waveform with $R=0.1$ versus the frequency. Bottom panel: amplitude of the 32 mode for the no-multibanding and multibanding. Notice the correspondence of the maximum phase error with the deep in the amplitude of the 32 mode.  }
    \label{fig:badcase_32}
\end{figure}

In Fig.~\ref{fig:badcase_32} we show some typical behaviour for mode mixing in the ringdown: Here the complex waveform comes very close to or crosses zero, visible as a sharp feature in the logarithm of the amplitude. Near the zero-crossing splitting the waveform into a spherical harmonic amplitude and phase creates artefacts when computing phase differences or relative amplitude errors between two waveforms, even if they are very close. Comparing our theoretical thresholds with the phase only makes sense in the spheroidal basis, but not in the spherical one. We omit a comparison of the phase errors in the ringdown as computed in the spheroidal picture in order to avoid excess baggage in our \texttt{LALSuite} implementation. We thus arrive at the following interpretation of Fig.~\ref{fig:error_phase_32}: while phase errors are typically small and below the threshold, a significant number of outliers arise due to the phenomenon shown in Fig.~\ref{fig:badcase_32}, they are however not due to problems of the multibanding algorithm, but due to keeping our test simple and uniformly comparing in the spherical harmonic picture for all modes and across the whole frequency range.

% Despite the maximum error may sometimes overpass the threshold, the mean error for all the frequency points is always below the threshold for all the configurations, therefore we consider that the threshold value is a suitable measure of the error that we are committing. Moreover the threshold value can also be used as a measure of the relative error in the total waveform that is introduce by the multibanding, see Fig.~\ref{fig:error_wf}. We compute the relative error between two waveforms for each frequency point as
% \begin{equation}
%     \mathrm{Rel\: Error} = \bigg\vert 1 - \frac{h_1(f)}{h_2(f)}\bigg\vert.
% \end{equation}
% The results are very similar to what we found in Fig.~\ref{fig:error_phase}, the maximum relative error may be above the corresponding threshold but the average relative error is practically always below. Note that the algorithm was not built such that the relative error satisfies this relation, but it is something that we get for free. 

% \begin{figure}[ht]
%     \centering
%     \includegraphics[width=\columnwidth]{"Plots/Max relative wf error".pdf}
%     \includegraphics[width=\columnwidth]{"Plots/Mean relative wf error".pdf}
%     \caption{Top panel: Maximum relative error for the lm mode between the multibanding and no-multibanding waveforms for four values of the threshold $R$ and 140000 configurations. Bottom panel: same than above but showing the mean relative error across the frequency array instead of the maximum.}
%     \label{fig:error_wf}
% \end{figure}

%%%%%%%%%%%%%%% Random Exploration %%%%%%%%%%%%%%%%%%%
To truly understand the accuracy of the algorithm we have to compute the mismatch between the two waveforms. In the following we evaluate the multimode waveform and compute the mismatch for the $h_+$ polarization. We carry out an extensive study across the whole parameter space also to test the robustness of the algorithm and evaluate one million of random configurations in the parameter space. 
The results are shown in Fig.~\ref{fig:mband_matches}. As expected the threshold $10^{-4}$ has the lowest mismatches since it is the most accurate and the threshold $0.1$ has the worst mismatches because it is the less accurate. The reader may be wondering why there are a significant number of cases with mismatch $10^{-16}$ since we would expect that the number of cases decreases with the higher accuracy. The explanation is that in this bin the randomly chosen $df$ is coarser than the $\Delta f$ that the multibanding criteria provides, and in this case we replace $\Delta f$ with $df$ and in consequence there is no difference between the multibanding and no-multibanding and we reach matching precision. 
%It might be striking that the $R=10^{-4}$ threshold has more points for the mismatch $10^{-16}$ than for others mismatches, this happens because as we mentioned earlier, this threshold can be so restrictive that the spacing of the coarse frequency may be smaller than the spacing of the fine grid $df$ that the user specifies, therefore we just use the input $df$ and then the two waveforms agree up to machine precision that is $10^{-16}$. 

Fig.~\ref{fig:mband_matches} is also very useful to decide which threshold we want to use as default value in the \texttt{LAL} code. We consider that $R=10^{-3}$ performs well in accuracy and given that is faster than the $10^{-4}$ we set this as the default value. 

%%%%%%%%%%%%%%%%%%%%%%%%%% PARAMETER ESTIMATION %%%%%%%%%%%%%%%%%%%%%%
\subsection{Parameter Estimation}\label{sec:pe}

In order to illustrate our algorithm in a parameter estimation application, we compare the performance of the original model and different choices for the threshold parameter $R$.

We select a publicly available numerical relativity data set from the SXS waveform catalogue \cite{SXS:catalog},
{\tt SXS:BBH:0264}, which corresponds to a binary black hole merger at mass ratio 3, with individual spins of $-0.6$ anti-aligned with the orbital momentum. Then we inject this numerical relativity simulation into zero noise as a way to get a non-precessing and non-eccentric strain of 4 seconds of duration, with $100~M_{\odot}$ total mass, near edge on with $\pi/3$ rad of inclination. We use a relatively close source at 430 Mpc, which implies a signal-to-noise ratio of 28. Recovery of the signal uses the advanced LIGO zero detuning high power noise curve \cite{LIGOPSD}.
We choose the parameters in order to challenge our approximations in the regime where higher modes are particularly relevant, not in order to demonstrate significant computational gains, which by Fig.~\ref{fig:T_vs_M_q} when the lower cutoff frequency of the detector sensitivity would be lower than the 20 Hz we have chosen here to compute the likelihood function in our Bayesian inference algorithm (see e.g.~\cite{Veitch:2014wba,Ashton:2018jfp} for details of Bayesian inference for compact binary coalescence signals). Note that the start frequency of the numerical relativity waveform we choose here is approximately 9 Hz at $M=100 M_\odot$.

For our analysis we use a sampling method called ``Nested Sampling'' \cite{Skilling:2006gxv}, in particular the CPNest sampler \cite{JohnVeitchCpnest} as implemented in the Python-based Bayesian inference framework Bilby \cite{Ashton:2018jfp}. 
For each waveform model used, we carry out runs with five different seeds and 2048 ``live points'' in the language of nested sampling, and we merge the results from the five seeds to a single posterior result.

We define prior distributions as follows: The mass ratio is assumed to be uniform between $0.125$ and $1$, and the chirp mass prior is assumed uniform between 15 and $60~M_{\odot}$. The luminosity distance is uniform in volume with a
maximal allowed distance at 1500 MPc. Finally, the magnitudes of the dimensionless black hole spins are uniform with an upper limit at  $0.99$.

Our main results concern the comparison of the \phXHM model, evaluated with different values of the threshold parameter, $R=(10^{-1},~10^{-2}, ~10^{-3})$ as well as without multibanding, which corresponds to $R=0$.
The  \phXHM model includes the spherical harmonic modes $(l,\left|m\right|) = (2,2),(2,1),(3,2),(3,3),(4,4)$. Differences between the recovered value of parameters and the injected parameters may arise due to the approximations in our multibanding algorithm, errors in the \phXHM model, errors in the numerical relativity waveform, and the absence of modes in the model, which are present in the numerical relativity data set (which contains all modes up to $l=8$). We also compare with the \phX model, which corresponds to \phXHM with only the $(l,\left|m\right|) = (2,2)$ modes and no multibanding. 
The latter serves as a comparison in terms of the errors in recovering the injection parameters.

\begin{figure*}[htpb]
    \centering
\includegraphics[width=\columnwidth]{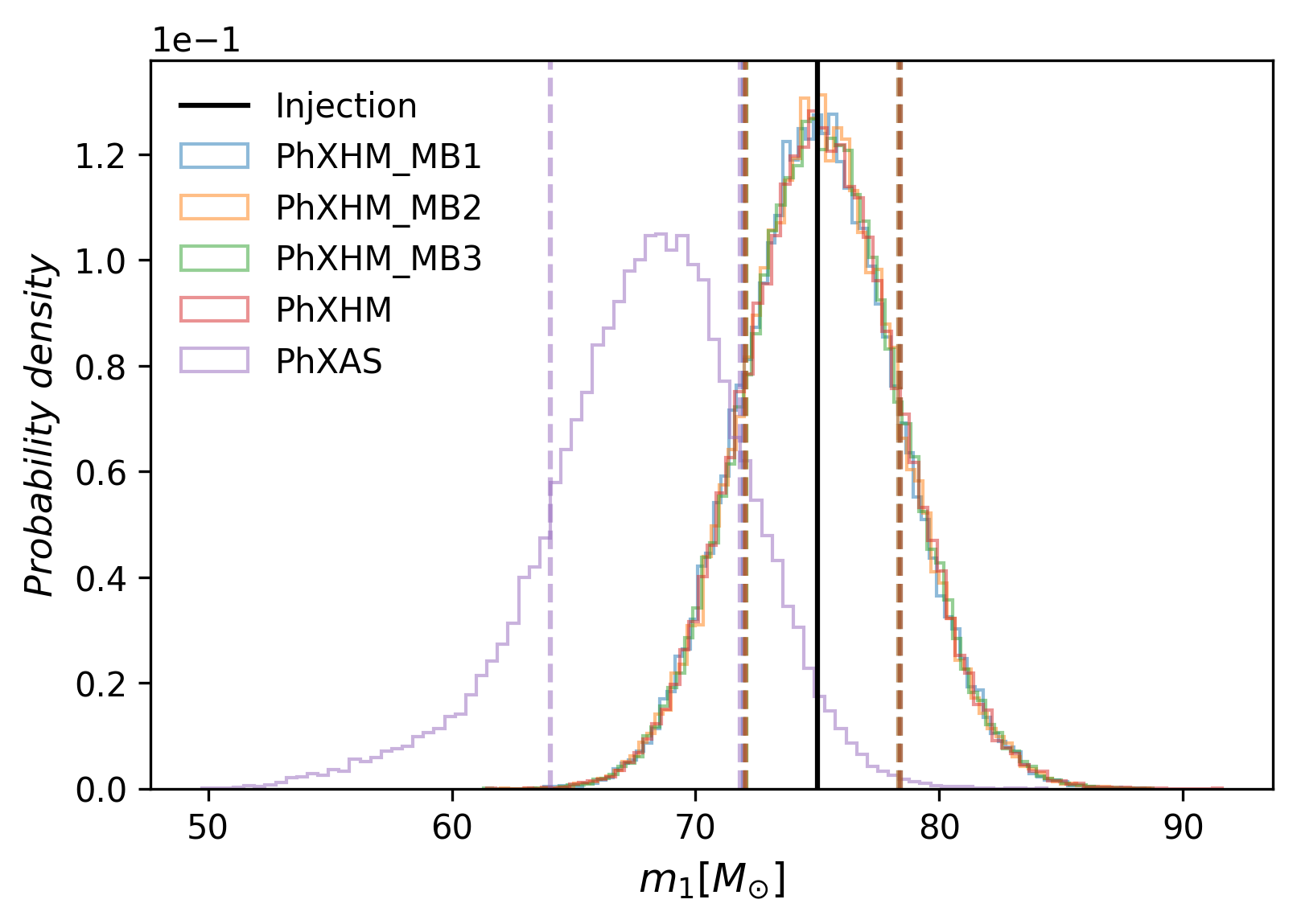}\includegraphics[width=\columnwidth]{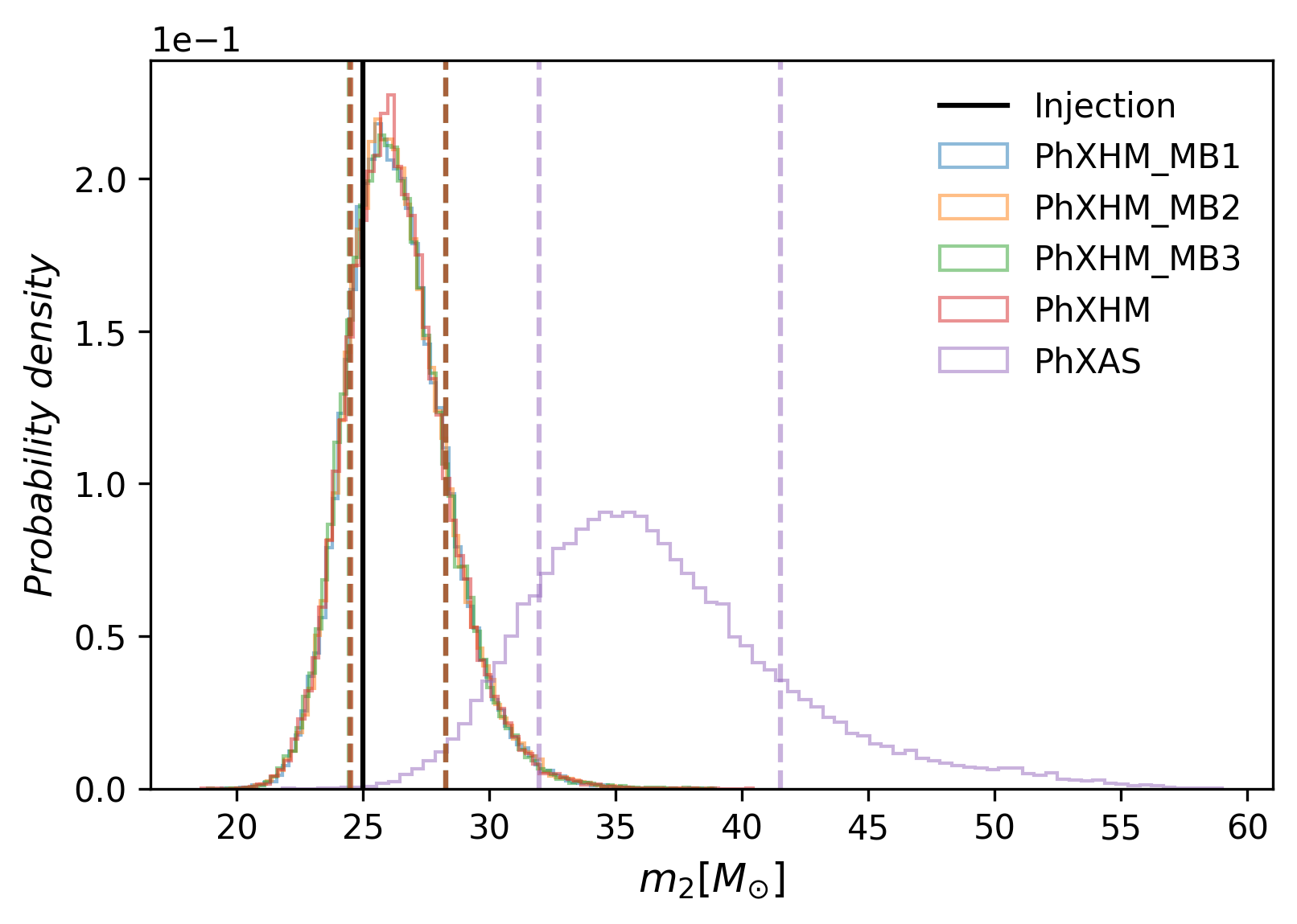}
\includegraphics[width=\columnwidth]{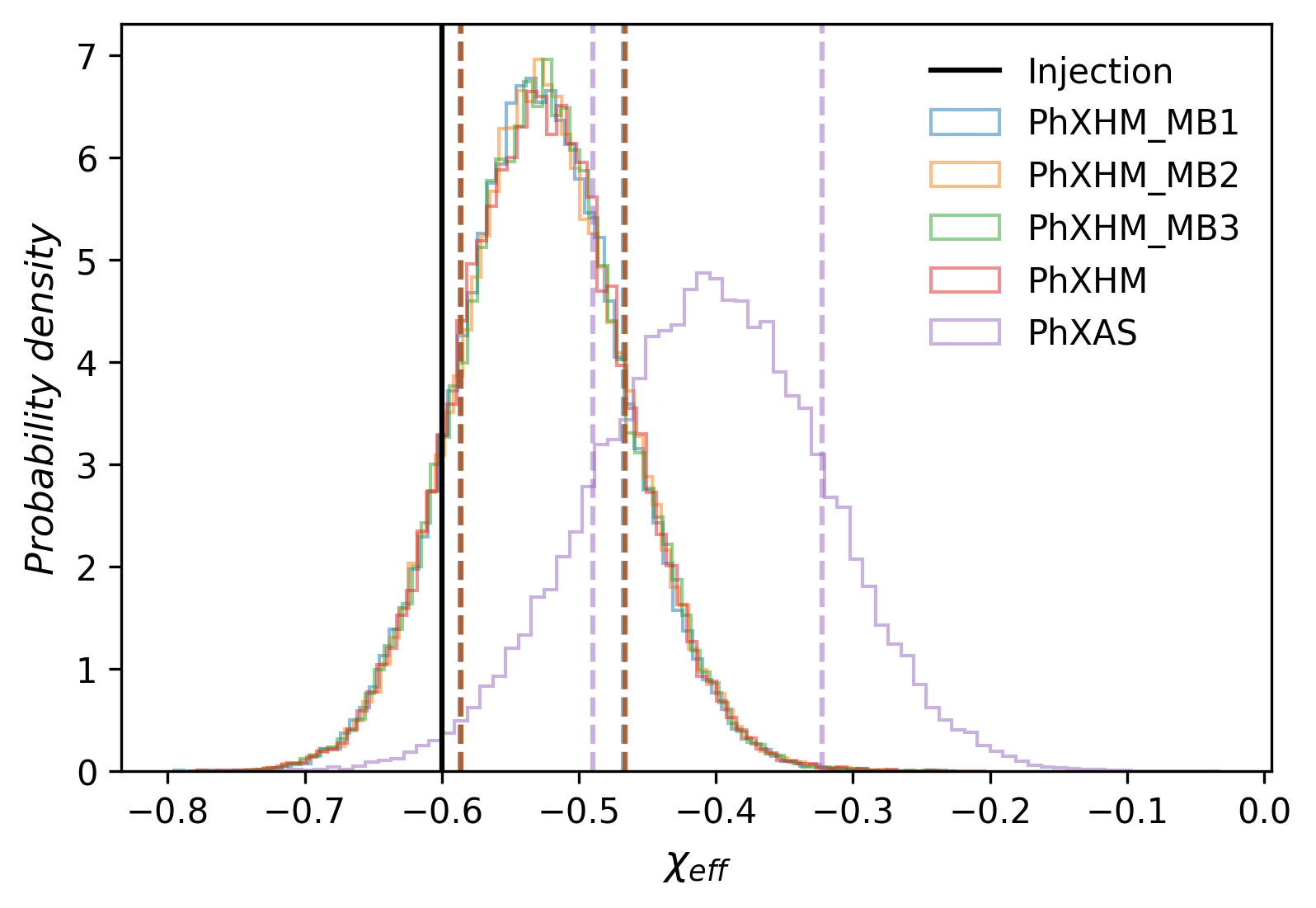}\includegraphics[width=\columnwidth]{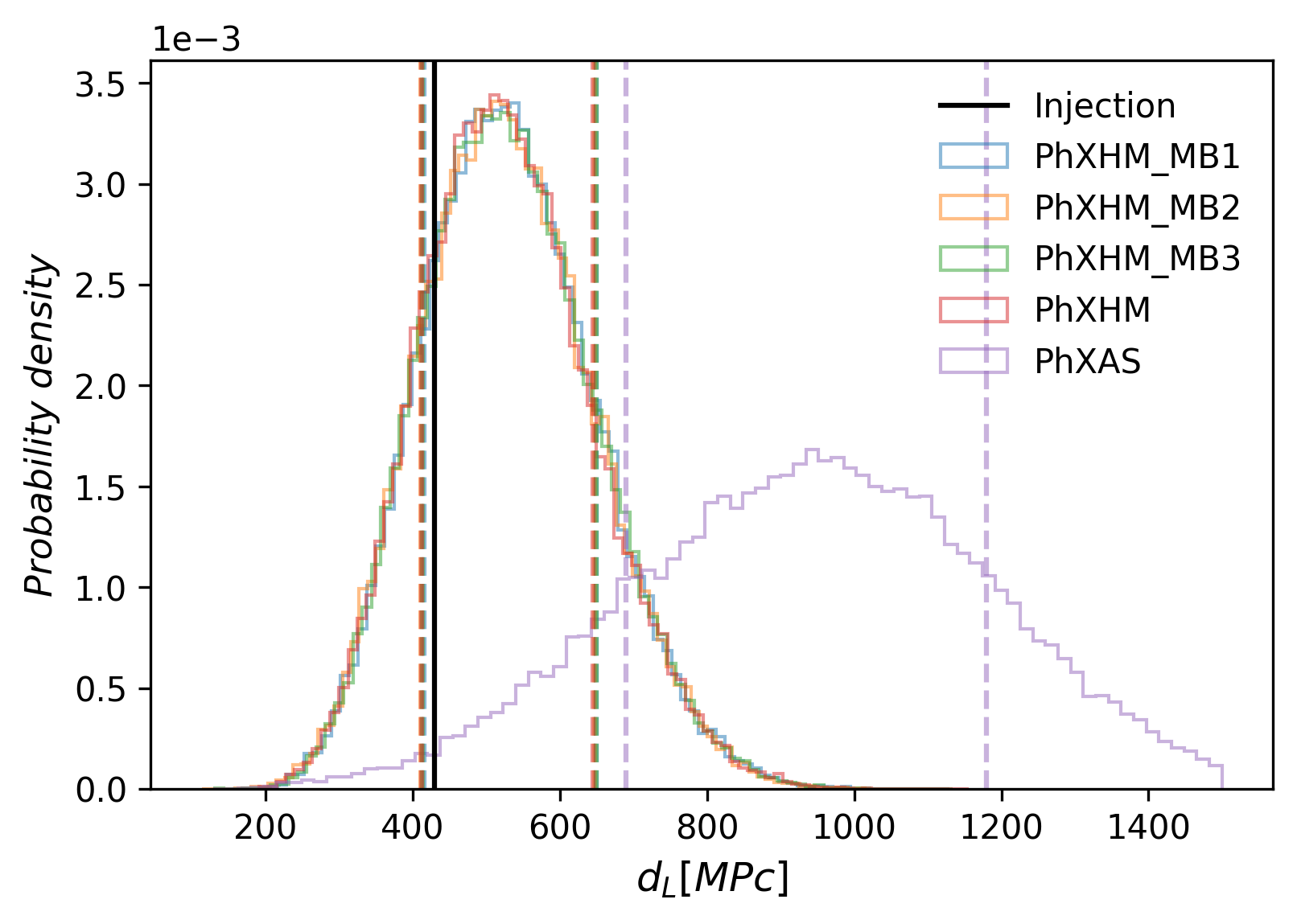}
    \caption{Posterior distributions of component masses, effective aligned spin and inclination respectively using waveforms with multibanding (PhXHM\_MB) for different values of threshold ($10^{-1}$, $10^{-2}$, $10^{-3}$) and without it (PhXHM and PhXAS). The dashed vertical lines mark the $90\%$ confidence limits.}
    \label{fig:PE}
\end{figure*}

Our results are presented in Fig.~\ref{fig:PE}.
In the case of the higher modes model, the injected values are recovered by the most probable regions of the posterior distributions. However, for the dominant mode model, a significant bias in the recovered parameters can be observed.
This confirms the importance of the higher mode contributions for the case we have chosen.
All the results for \phXHM are consistent within the statistical errors implied by our finite sampling. 
As expected, in the case presented here the sampling time only decreases weakly when increasing the threshold value. 
We attribute the observed parameter bias for \phXHM to the incomplete set of modes described by the model, as well as modelling errors. Future work will investigate the effect of dropping modes in the model in more detail.

%%%%%%%%%%%%%%%%%%%%%%%%%%%%%%%%%%%%%%%%%%%%%%%%%%%%%%%%
\section{Conclusions} \label{sec:conclusions} 

We have presented a simple way to accelerate the evaluation of frequency domain
waveforms by first evaluating on a coarse grid, and then interpolating to a fine grid with an iterative scheme to evaluate complex exponential functions (or equivalently trigonometric functions).
This works builds upon the method presented in \cite{Vinciguerra:2017ngf},
but represents the heuristic criterion used there to determine the spacing of the coarse grid by the standard estimate for first order interpolation error, and then extends the criterion for the coarse frequency spacing to the merger and ringdown.
Several extensions of our algorithm are possible: First, similar techniques can also be developed for the time domain. The simple estimates to determine the appropriate coarse grid spacing given a threshold parameter could be improved, e.g.~by adding low order spin terms. The amplitude could be treated in a similarly careful way as the phase.

Acceleration is more significant for smaller spacings of the fine grid, as is appropriate for smaller masses, and for detectors with broader sensitivity in frequency, e.g. future detectors such as the upgrades of the current generation of the advanced detector network, the Einstein Telescope \cite{ET} or LISA \cite{LISA}. For total masses around three solar masses, as is appropriate for binary neutron star masses, the current speed of  the multi-mode \phXHM roughly equals the speed of \phX for the $\ell=\vert m\vert$ modes. Detailed profiling of the code reveals this rough equality as a coincidence, and performance is limited by a small number of bottlenecks, e.g.~evaluating the spline interpolation for the amplitude, for which we use the \texttt{GSL} library \cite{gsl}. Future optimization work will focus on these bottlenecks.
Another possible avenue for further speedup would be an implementation on GPUs or similar highly parallel hardware.

The availability of a threshold parameter that regulates accuracy and speed also allows future applications to tune codes for parameter estimation, where the threshold parameter could be set depending on the information associated with the detection in a search (such as the signals rough parameter estimate from the search and its signal-to-noise ratio), or the threshold could be changed dynamically, and could be relaxed in the burn-in-phase of a parameter estimation simulation, or in the early stages of a nested sampling run. 
The coupling of strategies to accelerate the evaluation of individual waveforms evaluation and Bayesian parameter estimation simulations as a whole may also have implications on the development of future waveform models, which could introduce further parameters to tune accuracy and evaluation speed.

%%%%%%%%%%%%%%%%%%%%%%%%%%%%%%%%%%%%%%%%%%%%%%%%%%%%%%%%

\section*{Acknowledgements} 
We thank G.~Pratten, M.~Colleoni for discussions on the development of the \phX and \phXHM \cite{phenX,phenXHM} models, and the internal reviewers of the LIGO-Virgo collaboration for their careful checking of our \texttt{LALSuite} code implementation and their valuable feedback.
This work was supported by European Union FEDER funds, the Ministry of Science, 
Innovation and Universities and the Spanish Agencia Estatal de Investigación grants FPA2016-76821-P,        % FPA2017-90687-REDC, FPA2017-90566-REDC. 
RED2018-102661-T,    % RENATA
RED2018-102573-E,    % REDES ESTRATÉGICAS: Participación Española en Estructuras Euro... 
FPA2017-90687-REDC,  % CPAN
Vicepresid`encia i Conselleria d’Innovació, Recerca i Turisme, Conselleria d’Educació, i Universitats del Govern de les Illes Balears i Fons Social Europeu, 
Generalitat Valenciana (PROMETEO/2019/071),  
EU COST Actions CA18108, CA17137, CA16214, and CA16104, and
the Spanish Ministry of Education, Culture and Sport grants FPU15/03344 and FPU15/01319.
%
%MC acknowledges funding from the European Union's Horizon 2020 research and innovation programme, under the Marie Skłodowska-Curie grant agreement No. 751492.
%
The authors thankfully acknowledge the computer resources at MareNostrum and the technical support provided by Barcelona Supercomputing Center (BSC) through Grants No. AECT-2019-2-0010, AECT-2019-1-0022, AECT-2018-3-0017, AECT-2018-2-0022, AECT-2018-1-0009, AECT-2017-3-0013, AECT-2017-2-0017, AECT-2017-1-0017, AECT-2016-3-0014, AECT2016-2-0009,  from the Red Española de Supercomputación (RES) and PRACE (Grant No. 2015133131). {\tt Bilby} and \texttt{LALInference} simulations were carried out on the BSC MareNostrum computer under RES (Red Española de Supercomputación) allocations and on the FONER computer at the University of the Balearic Islands. The authors are also grateful for computational resources provided by the LIGO Laboratory and supported by National Science Foundation Grants PHY-0757058 and PHY-0823459.

%%%%%%%%%%%%%%%%%%%% References %%%%%%%%%%%%%%%%%%%%%%%%%%%%%%%%%%%

\bibliography{uib,phenomx}

\end{document}